\documentclass[journal,a4paper]{IEEEtran}

\usepackage[hidelinks]{hyperref}

\newif\ifonecolumn
\usepackage{cite}

\ifCLASSINFOpdf
  \usepackage[pdftex]{graphicx}
\else
\fi

\usepackage{amsmath}
\usepackage{amsthm}
\usepackage{amssymb}
\usepackage{array}
\usepackage{booktabs} % nice rulers for tables

\ifCLASSOPTIONcompsoc
  \usepackage[caption=false,font=normalsize,labelfont=sf,textfont=sf]{subfig}
\else
  \usepackage[caption=false,font=footnotesize]{subfig}
\fi

\usepackage{fixltx2e}

\usepackage{lipsum}
\usepackage{datetime}    % to get the \currenttime command
\usepackage{multirow}	% to have multirows
\usepackage{nicefrac}   % nice fractions with "/"
\usepackage{color}
\usepackage{bm}		% bold greek letters in formulas
\usepackage{enumitem}  % i), ii), ... for enumerations
\usepackage{balance} % to balance two columns on the last page
\usepackage[np]{numprint} % thousand separators in numbers
	\npdecimalsign{.}
	\npfourdigitnosep

\usepackage{import}
\newtheorem{thm}{Theorem}
\newtheorem{lemma}{Lemma}
\newtheorem{cor}{Corollary}
\newtheorem{define}{Definition}
\newtheorem{note}{Note}

\usepackage[capitalize]{cleveref}
  \crefname{equation}{}{}
  \crefname{thm}{Theorem}{Theorems}
  \crefname{lemma}{Lemma}{Lemmas}
  \crefname{cor}{Corollary}{Corollaries}
  \crefname{define}{Definition}{Definitions}
  \crefname{note}{Note}{Notes}
  \crefname{propose}{Proposition}{Propositions}
  \crefname{appsec}{Appendix}{Appendices}

\newcommand{\code}{\mathcal C}
\newcommand{\dual}{\code^\perp}
\newcommand{\ensemble}{\mathfrak C}
\renewcommand{\vec}[1]{\bm{#1}}
\newcommand{\cE}{\mathcal{E}}
\newcommand{\cS}{\mathcal{S}}

\newcommand{\bbN}{\mathbb{N}}

\newcommand{\Prob}[2][]{\mathbb P_{#1} \left\{ #2 \right\}}
\newcommand{\Expect}[2][]{\mathbb E_{#1} \left\{ #2 \right\}}
\DeclareMathOperator{\rank}{rank}

\DeclareMathOperator{\supp}{supp}

\usepackage[capitalize]{cleveref}
	\crefname{equation}{}{}
	\crefname{thm}{Theorem}{Theorems}

\begin{document}
\title{Stopping Redundancy Hierarchy Beyond the Minimum Distance}
\author{Yauhen~Yakimenka,~\IEEEmembership{Student~Member,~IEEE, $\;$}
        Vitaly~Skachek,~\IEEEmembership{Member,~IEEE,}
        Irina~E.~Bocharova,~\IEEEmembership{Senior~Member,~IEEE,$\;$}
        Boris~D.~Kudryashov,~\IEEEmembership{Senior~Member,~IEEE}% <-this % stops a space
\thanks{Y. Yakimenka, V. Skachek, I.E. Bocharova and B.D. Kudryashov are with the Institute of Computer Science, University of Tartu, Estonia. I.E. Bocharova and B.D. Kudryashov are also with ITMO, St. Petersburg, Russian Federation.}%
\thanks{The results of this work were partly presented in~\cite{yakimenka-skachekITW}.}
\thanks{This work was supported in part by the Norwegian-Estonian Research Cooperation Programme grant EMP133, and the Estonian Research Council grants PUT405 and PRG49. Calculations were partially carried out in the High Performance Computing Centre of the University of Tartu.}
\thanks{Copyright (c) 2018 IEEE. Personal use of this material is permitted.  However, permission to use this material for any other purposes must be obtained from the IEEE by sending a request to pubs-permissions@ieee.org.}}

% If you want to put a publisher's ID mark on the page you can do it like
% this:
%\IEEEpubid{0000--0000/00\$00.00~\copyright~2015 IEEE}
% Remember, if you use this you must call \IEEEpubidadjcol in the second
% column for its text to clear the IEEEpubid mark.

% use for special paper notices
%\IEEEspecialpapernotice{(Invited Paper)}

% make the title area
\maketitle

% As a general rule, do not put math, special symbols or citations
% in the abstract or keywords.
\begin{abstract}
Stopping sets play a crucial role in failure events of iterative decoders over a binary erasure channel (BEC). The $\ell$-th stopping redundancy is the minimum number of rows in the parity-check matrix of a code, which contains no stopping sets of size up to $\ell$. In this work, a notion of coverable stopping sets is defined. In order to achieve maximum-likelihood performance under iterative decoding over the BEC, the parity-check matrix should contain no coverable stopping sets of size $\ell$, for $1 \le \ell \le n-k$, where $n$ is the code length, $k$ is the code dimension. By estimating the number of coverable stopping sets, we obtain upper bounds on the $\ell$-th stopping redundancy, $1 \le \ell \le n-k$. The bounds are derived for both specific codes and code ensembles. In the range $1 \le \ell \le d-1$, for specific codes, the new bounds improve on the results in the literature. Numerical calculations are also presented. 
\end{abstract}

% Note that keywords are not normally used for peerreview papers.

% For peer review papers, you can put extra information on the cover
% page as needed:
% \ifCLASSOPTIONpeerreview
% \begin{center} \bfseries EDICS Category: 3-BBND \end{center}
% \fi
%
% For peerreview papers, this IEEEtran command inserts a page break and
% creates the second title. It will be ignored for other modes.
\IEEEpeerreviewmaketitle

\section{Introduction}
\IEEEPARstart{L}{ow}-density parity-check (LDPC) codes~\cite{gallager1962low} in conjunction  with iterative (message-passing, MP) decoding is a popular choice for error correction in practical communications and data storage systems. They allow for very fast decoding, while the decoding performance is suboptimal. 

Recently, a number of new applications, in particular, in the domain of network communications and data storage, have raised interest in communications over the binary erasure channel (BEC). LDPC codes can be used over the BEC. The maximum-likelihood (ML) decoder for codes used over the BEC is equivalent to solving a sparse system of linear equations, which is relatively time-consuming for practical applications. An alternative decoding method based on iterative procedure offers complexity linear in the code length $n$, yet its decoding error performance is suboptimal. 

On the BEC, the MP decoding algorithm can be interpreted as an iterative process to repetitively solve a single equation with one unknown. By contrast, on this channel, ML decoding is equivalent to solving a sparse system of linear equations, which can be done in $O(n^3)$ operations by using Gaussian elimination. More elaborate algorithms allow for solving a sparse system of equations in time approximately $O(n^2)$ (cf. \cite{wiedemann1986, mofrad2013}), which is still too high for many practical applications.

It was observed in~\cite{di2002finite} that \emph{stopping sets} are combinatorial structures in a parity-check matrix of a code, which are responsible for failures of iterative decoding on the BEC. Thus, stopping sets in the parity-check matrix should be eliminated in order to decrease the failure rate of the decoder. 

In~\cite{schwartz-vardy2006}, it was suggested to adjoin redundant rows to a parity-check matrix of a code in order to eliminate stopping sets of size less than $d$, where $d$ is the minimum distance of a code. It was shown that this can be done for any code by taking a sufficiently large number of redundant rows of the parity-check matrix. Bounds on the \emph{stopping redundancy}, the minimum number of the redundant rows, were studied in \cite{schwartz-vardy2006} and subsequently improved in~\cite{han2007improved,han2008improved}.   

In~\cite{olgica2008permutation}, the authors defined the so-called \emph{stopping redundancy hierarchy}. In their approach, the $\ell$-th stopping redundancy is the minimum number of rows in the parity-check matrix, such that the parity-check matrix contains no stopping sets of size at most $\ell$. Algorithms for finding small stopping sets were proposed in~\cite{rosnes2009efficient, karimi2013efficient}. 

Distribution of the number of stopping sets of different sizes in the Hamming code was studied in~\cite{weber2005stopping}. Other related papers include, for example,~\cite{kashyap2003, etzion2006stopping, hollmann2006generic, hollmann2006generatingA, hollmann2006generatingB, hollmann2007parity, zumbraegel2012pseudocodeword}. 

In this work, we first propose an improvement to the upper bounds in~\cite{han2007improved, han2008improved}. While the upper bounds in~\cite{han2007improved, han2008improved} are obtained by probabilistic methods, in our approach, we initially select a few first rows deterministically and then continue along the lines of probabilistic analysis in~\cite{han2007improved,han2008improved}. 

Next, we consider stopping sets of size $\ell \ge d$. We define \emph{coverable} stopping sets, which do not contain a support of any nonzero codeword. We extend the definition of the stopping redundancy hierarchy beyond size $\ell = d-1$ of the stopping sets. By estimating the number of coverable stopping sets, we derive upper bounds on the stopping redundancy hierarchy also for $\ell \ge d-1$. Finally, the values of the bounds for specific codes and ensembles are calculated, and it is shown that theoretical finding are consistent with the experimental results.    

The structure of this paper is as follows. In \cref{sec:preliminaries}, the necessary notations are introduced and the basic facts used in the paper are established. \cref{sec:stop_redundancy} contains improved bounds on the stopping redundancy. In \cref{sec:ML-perf}, we study stopping sets of size $d$ and higher, and the bounds on the stopping redundancy hierarchy for $\ell \ge d$ are derived. Both specific codes and code ensembles are considered in that section. Further, we experimentally obtain numerical results in \cref{sec:simulations}. \cref{sec:conclusions} concludes the paper.

% needed in second column of first page if using \IEEEpubid
%\IEEEpubidadjcol

\section{Preliminaries}
\label{sec:preliminaries}
\subsection{Notations}
We start this section by introducing some notations used throughout the paper.  

Consider a binary linear $[n, k, d]$ code $\code$, where $n$, $k$, and $d$ denote length, dimension, and minimum distance, respectively. Let $H = (h_{ji})$ be a binary $m \times n$ parity-check matrix of this code, and $\rank H = r \triangleq n-k$ be the dimension of the dual code $\dual$. We note that generally $r \le m$.

We denote $[n] \triangleq \{ 1,2,\dotsc,n \}$. Let $\cS \subseteq [n]$ be a set of column indices of a matrix $H$. Denote by $H_{\cS}$ the submatrix of $H$ consisting of the columns of $H$ indexed by $\cS$. 

A \emph{support} of a vector $\vec{x}$, denoted by $\operatorname{supp}(\vec{x})$, is the set of indices of nonzero entries in the vector. \emph{Hamming weight} of a vector is a cardinality of its support.

\subsection{Stopping Sets and Stopping Redundancy}
We start this section by defining a stopping set.

\begin{define}[\hspace{-0.6ex} \cite{di2002finite}]
Let $H$ be an $m \times n$ parity-check matrix of a binary code $\code$. The set $\cS \subseteq [n]$ is called the \emph{stopping set} if $H_{\cS}$ contains no row of Hamming weight one. 
\end{define}

It is important to note that stopping sets are structures in a particular parity-check matrix, and not in the code.
%\smallskip

Following the terminology of \cite{schwartz-vardy2006}, we formulate the following definition.
\begin{define}
\label{def:cover} 
A binary vector $\vec h$ \emph{covers} a stopping set (or any subset of columns) $\cS$ if $\operatorname{supp}(\vec h)$ intersects with $\cS$ in exactly one position. Consequently, a matrix covers $\cS$ if any of its rows covers $\cS$. 
\end{define}

We note that if $\cS$ is a stopping set in a parity-check matrix $H$ and $\vec h$ covers $\cS$, then after adjoining $\vec h$ as a row to $H$, $\cS$ is not a stopping set in the obtained extended matrix. With some abuse of notation, we say that a stopping set $\cS$ is \emph{covered} in that extended matrix.\footnote{In other words, we will use ``the stopping set $\cS$ is covered by the matrix'' and ``$\cS$ is not a stopping set in the matrix'' interchangeably.}

\begin{define}
\label{def:coverable} 
A stopping set $\cS$ is \emph{coverable} (by the code $\code$), if there exists a (possibly extended) parity-check of $\code$ that covers $\cS$. 
\end{define}

In order to reduce the failure probability of the iterative decoding algorithm over the BEC, it was proposed in~\cite{schwartz-vardy2006} to adjoin redundant rows, i.e., the codewords of $\dual$, to a parity-check matrix in such a way that the resulting matrix has no stopping sets of small size. Specifically, we are interested in adjoining the minimum possible number of codewords from $\dual$ to a parity-check matrix $H$ such that all the stopping sets of size less than $d$ become covered. It was shown in \cite{schwartz-vardy2006} that this is always possible, and thus all stopping sets of size less than $d$ are coverable.
 
In this work, we build on the approach in~\cite{schwartz-vardy2006}. Namely, we extend parity-check matrices by choosing codewords from $\dual$ and adjoining them as redundant rows. The extended matrices are constructed in such a way that they do not contain stopping sets of small size. In the sequel, we provide a detailed analysis of the minimum number of additional rows in order to achieve this goal. In what follows, we use the terms ``row of a parity-check matrix'' and ``codeword from $\dual$'' interchangeably. We also note that a particular order of rows in a parity-check matrix is not important. As a matter of convenience, we will denote by $\dual_0$ the set of all the codewords of $\dual$ except the zero vector.

\begin{define}[\hspace{-0.6ex} \cite{schwartz-vardy2006}]
	The size of the smallest stopping set of a parity-check matrix $H$, denoted by $s(H)$, is called the \emph{stopping distance} of the matrix.
\end{define}

It is known that a maximal parity-check matrix $H^{(2^r)}$ consisting of all $2^r$ codewords of $\dual$ is an orthogonal array of strength $d-1$ (cf. \cite[Ch.~5, Thm.~8]{macwilliams1977theory}). This means that for any $\cS \subseteq [n]$ of size $i$, $1 \leq i \leq d-1$, $H_\cS^{(2^r)}$ contains each $i$-tuple as its row exactly $2^{r-i}$ times and, hence, $\mathcal S$ is covered by exactly $i \cdot 2^{r-i}$ rows of $H^{(2^r)}$. 

The following definition was introduced in~\cite{schwartz-vardy2006}. 
\begin{define}\label{def:stopping-redundancy}
The stopping redundancy of $\code$, denoted by $\rho(\code)$, is the smallest number of rows in any parity-check matrix (of rank $r$) of $\code$, such that the corresponding stopping distance equals $d$. 
\end{define}  

It was shown in \cite[Thm.~3]{schwartz-vardy2006}, that \emph{any} parity-check matrix $H$ of a binary linear code $\code$ with the minimum distance $d \leq 3$ already has $s(H) = d$. In what follows, we are mostly interested in the case $d > 3$.

%%%%%%%%%%%%%%%%%%%%%%%%%%%%%%%%%%%%%%%%%%%%%%%%%%%%%%%%%%%%%%%%%%%%%%%%%%%%%%%%%%%%%%%%%

\section{Upper Bounds on Stopping Redundancy}
\label{sec:stop_redundancy}
Before proceeding further, we present the following technical result. It will be used in the proof of \cref{thm:sr-itw}.

\begin{lemma}\label{lem:1minusfrac}
	For any integers $i,j,r \geq 1$, and $j < 2^r$, define
	\[
	\pi(r, i, j) \triangleq 1 - \frac{i \cdot 2^{r-i}}{2^r-j} .
	\]
	Then, for any integer $r \leq r'$, and $i \leq i'$, we have $\pi(r,i,j) \leq \pi(r',i,j)$ and $\pi(r,i,j) \leq \pi(r,i',j)$.	In other words, $\pi(r,i,j)$ is monotonically non-decreasing in integer variables $r$ and $i$.
\end{lemma}
\begin{IEEEproof}
%	The lemma is proven by taking partial derivatives (i.e., considering $\pi(\cdot)$ as a function of real parameters). First, for any values in the aforementioned region, 
%	\[
%	\frac{\partial}{\partial r} \left( 1 - \frac{i \cdot 2^{r-i}}{2^r - j} \right) = \frac{i j \cdot \log (2) \cdot 2^{r-i}}{\left(2^r-j\right)^2} > 0.
%	\]
%	Second, for $i \geq 2 > 1/\log(2)$ (and $j < 2^r$),
%	\[
%	\frac{\partial}{\partial i} \left( 1 - \frac{i \cdot 2^{r-i}}{2^r - j} \right) = \frac{2^{r-i} (i \cdot \log (2)-1)}{2^r-j} > 0.
%	\]
%	To complete the proof, we remark that $\pi(r,1,j) = \pi(r,2,j)$.
The statement of the lemma follows easily if we rewrite:
	\[
	\pi(r,i,j) = 1 - \frac{i}{2^i} \cdot \frac{1}{1-j \cdot 2^{-r}}.\qedhere
	\]
\end{IEEEproof}

\subsection{Upper Bounds for General Codes}
In \cite{schwartz-vardy2006}, Schwartz and Vardy presented the first upper bound on the stopping redundancy of a general binary linear $[n,k,d]$ code $\code$:
\begin{equation}\label{eq:SV-bound}
\rho(\code) \leq \binom r1 + \binom r2 + \dotsb + \binom{r}{d-2}.
\end{equation}
This bound is constructive. The other works along the same lines are \cite{weber2005stopping,hollmann2006generatingA,hollmann2006generatingB,hollmann2006generic,hollmann2007parity,etzion2006stopping,han2007improved,olgica2008permutation}, which present other constructive upper bounds, either for general linear codes, or for some specific families, or for specific codes.

On the other hand, \textit{probabilistic} arguments gave rise to better bounds \cite{hollmann2006generatingB,hollmann2007parity,han2007improved,han2008improved,yakimenka-skachekITW}, yet these bounds are non-constructive. The main probabilistic technique in our paper dates back to the work of Han and Siegel \cite{han2007improved}:
\begin{equation}\label{eq:HS-bound}
\rho(\code) \leq \min \{ t \in \mathbb N \,:\, \mathcal E_{n,d}(t) < 1 \} + (r-d+1) \; ,
\end{equation}
where
\[
\mathcal E_{n,d}(t) \triangleq \sum_{i=1}^{d-1} \binom ni \left( 1 - \frac{i}{2^i} \right)^t \; .
\]
The bound has been improved in \cite{han2008improved} and further refined in \cite{yakimenka-skachekITW} by carefully selecting the first non-random rows, giving the smallest known values for most of codes (to the best of our knowledge).

Below we present a modified bound based on \cite[Thm. 1]{yakimenka-skachekITW}. More precisely, we drop the burdensome requirement 
\begin{equation}\label{eq:burdensome-req}
(r-1)(d-1) \leq 2^{d-1}
\end{equation}
thus making the bound applicable to all the binary linear codes. On the other hand, we need to add the rank deficiency term $\Delta$ to ensure that the constructed parity-check matrix has the required rank. However, for medium and long codes, this term is negligible in comparison with the stopping redundancy.

\begin{thm}\label{thm:sr-itw}
	For an $[n,k,d]$ linear binary code $\code$ let $H^{(\tau)}$ be any $\tau \times n$ matrix consisting of $\tau$ different codewords of the dual code $\dual$ and let $u_i$ denote the number of stopping sets of size $i$, $i=1,2,\dotsc,d-1$, in $H^{(\tau)}$. For $t = 0, 1, \cdots, 2^r-\tau$, we introduce the following notations: 
	\begin{gather*}
			\mathcal D_t = \sum_{i=1}^{d-1} u_i \prod_{j=\tau+1}^{\tau+t} \pi(r,i,j) , \\
			P_{t,0} = \lfloor \mathcal D_t \rfloor , \\
			P_{t,j} = \Big\lfloor \pi(r,d-1,\tau+t+j) \cdot P_{t,j-1} \Big\rfloor , \quad j=1,2,\dotsc \\
			\Delta = r - \max \{ \rank H^{(\tau)}, d-1 \} ,
	\end{gather*}
	and let $\kappa_t$
	be the smallest $j$ such that $P_{t,j} = 0$. Then
	\begin{equation}\label{eq:ub-ITW}
	\rho \leq \tau + \min_{0 \leq t < 2^r-\tau} \left\{ t + \kappa_t \right\} + \Delta .
	\end{equation}
\end{thm}
\begin{IEEEproof}
	We prove the theorem in two steps. First, we show the existence of a $(\tau + t) \times n$ matrix with a number of stopping sets less than or equal to $P_{t,0}$. Second, we show that this number further decreases when we add carefully selected rows one by one. Finally, after adding a sufficient number of rows, we obtain a matrix with no stopping sets of size less than $d$.
	
	\emph{Step 1}. By orthogonal array property, for any subset of columns $\cS \subseteq [n]$ of size $i$, $i = 1, 2, \dotsc, d-1$, there are exactly $i \cdot 2^{r-i}$ codewords in $\dual_0$, that cover $\cS$. If $\cS$ is not covered by $H^{(\tau)}$, none of these $i \cdot 2^{r-i}$ codewords is present among the rows of $H^{(\tau)}$.
	
	Fix a stopping set $\cS$ in $H^{(\tau)}$ and draw $t$ codewords from the set $\dual_0 \setminus \{ \textrm{rows of } H^{(\tau)} \}$ at random without repetition. There are 
	\[
	\binom{2^r - \tau - 1}{t}
	\]
	ways to do this, provided the order of selection does not matter. On the other hand, in the same set $\dual_0 \setminus \{ \textrm{rows of } H^{(\tau)} \}$, there are $(2^r-\tau-1)-i \cdot 2^{r-i}$ codewords that do \emph{not} cover $\cS$ and there are
	\[
	\binom{(2^r - \tau - 1) - i \cdot 2^{r-i}}{t}
	\]
	ways to draw $t$ codewords out of them. Therefore, if we draw $t$ codewords from the set $\dual_0 \setminus \{ \textrm{rows of } H^{(\tau)} \}$ at random without repetition, the probability not to cover $\cS$ by any one of them is (see \cref{app:binoms-divide-lemma})
	\ifonecolumn
	\begin{equation}\label{eq:binoms-divide-for-lemma}
	\left.
	\binom{(2^r - \tau - 1) - i \cdot 2^{r-i}}{t}
	\right/
	{\binom{2^r - \tau - 1}{t}}\\
	\overset{\scriptsize\text{\cref{lem:frac-binoms}}}{=} \prod_{j=\tau+1}^{\tau+t} \pi(r,i,j).
	\end{equation}
	\else
	\begin{multline}\label{eq:binoms-divide-for-lemma}
	\left.
	\binom{(2^r - \tau - 1) - i \cdot 2^{r-i}}{t}
	\right/
	{\binom{2^r - \tau - 1}{t}}\\
	\overset{\scriptsize\text{\cref{lem:frac-binoms}}}{=} \prod_{j=\tau+1}^{\tau+t} \pi(r,i,j).
	\end{multline}
	\fi	
	
	This holds for each $\cS$ that was not originally covered by $H^{(\tau)}$. Since numbers of stopping sets of sizes $1,2,\dotsc,d-1$ are $u_1,u_2,\dotsc,u_{d-1}$, respectively, the average\footnote{Averaging is by the choice of $t$ rows.} number of stopping sets of size less than $d$ that are left after adjoining random $t$ rows to $H^{(\tau)}$ is
	\begin{equation}
		\sum_{i=1}^{d-1} u_i \prod_{j=\tau+1}^{\tau+t} \pi(r,i,j)  \triangleq \mathcal D_t .
	\end{equation}
	
	Furthermore, since the above expression is an expected value of an integer random variable, there exists its realization (i.e., choice of $t$ rows), such that the number of stopping sets left is not more than $\lfloor \mathcal D_t \rfloor \triangleq P_{t,0}$. Fix these $t$ rows and further assume that we have a $(\tau+t) \times n$ matrix $H^{(\tau+t)}$ with not more than $P_{t,0}$ stopping sets of size less than $d$.
	
	\emph{Step 2}. Adjoin to $H^{(\tau+t)}$ a random codeword from $\dual_0 \setminus \{ \textrm{rows of } H^{(\tau+t)} \}$. If some stopping set $\cS$ of size $i$, $1 \le i \le d-1$, has not been covered by $H^{(\tau+t)}$ yet, there are exactly $i \cdot 2^{r-i}$ codewords in $\dual_0 \setminus \{ \textrm{rows of } H^{(\tau+t)} \}$ that cover $\cS$ and, thus, the probability that $\cS$ stays non-covered after adjoining this new row is
	\ifonecolumn
	\begin{equation*}
	1 - \frac{i \cdot 2^{r-i}}{2^r - (\tau+t+j)} = \pi(r,i,\tau+t+1)\\
	\overset{\scriptsize\text{\cref{lem:1minusfrac}}}{\leq}
	\pi(r,d-1,\tau+t+1).
	\end{equation*}
	\else
	\begin{multline*}
		1 - \frac{i \cdot 2^{r-i}}{2^r - (\tau+t+j)} = \pi(r,i,\tau+t+1)\\
		\overset{\scriptsize\text{\cref{lem:1minusfrac}}}{\leq}
		\pi(r,d-1,\tau+t+1).
	\end{multline*}
	\fi
	
	This holds for any stopping set $\cS$ of size $i$. Then, there exists a codeword in $\dual_0 \setminus \{ \textrm{rows of } H^{(\tau+t)} \}$ such that after adjoining it as a row to $H^{(\tau+t)}$, the number of non-covered stopping sets becomes less than or equal to 
	\[
	\Big\lfloor \pi(r,d-1,\tau+t+1) P_{t,0} \Big\rfloor \triangleq P_{t,1}.
	\]
	
	To this end, we fix this new row and further assume that we have a $(\tau+t+1) \times n$ matrix $H^{(\tau+t+1)}$ with the number of stopping sets of size smaller than $d$ less than or equal to $P_{t,1}$. After that, we iteratively repeat Step 2. We stop when the number of non-covered stopping sets is equal to zero.
	
	Finally, we need to ensure that the rank of the resulting matrix is indeed $r$. We already know that it is not less than $\rank H^{(\tau)}$. On the other hand, since we covered all the stopping sets of size less than $d$, the rank is at least $d-1$. Hence it is enough to add $\Delta$ additional rows to ensure the correct rank of the parity-check matrix.
\end{IEEEproof}

\begin{note}
	The expression in \cref{eq:ub-ITW} is monotonically non-decreasing in $u_i$. Sometimes, the exact values of $u_i$ are difficult to find. In that case, upper bounds are used instead. 
\end{note}

%	Unless it is stated otherwise, in this paper one can always use upper bounds on $u_i$ if the exact values cannot be calculated.

\begin{note}
	By applying \cref{lem:1minusfrac} to expressions for $\mathcal D_t$ and $P_{t,j}$, we obtain that \cref{eq:ub-ITW} is also monotonically non-decreasing in $r$. Sometimes, a standard parity-check matrix is redundant,\footnote{For instance, for Gallager $(J,K)$-regular codes (cf. \cref{sec:Gallager-average}).} and the number of rows $m$ in a mostly-used parity-check matrix is larger than $r$.  It might be more convenient to use $m$ instead of $r$ and the bound \cref{eq:ub-ITW} still holds.
\end{note}

To give a flavor of differences between the existing bounds on stopping redundancy, we calculate the bounds \cref{eq:SV-bound} in \cite{schwartz-vardy2006}, \cref{eq:HS-bound} in \cite{han2007improved}, the bound in \cite[Thm.~7]{han2008improved}, the bound in \cite[Thm.~1]{yakimenka-skachekITW}, and the bound in \cref{thm:sr-itw}. The two last bounds are calculated in two modes. First, we use $\tau=1$ and $H^{(\tau)}$ consists of the first row of the parity-check matrix of the corresponding code. Next, we use whole parity-check matrices of the codes as $H^{(\tau)}$ (in \cref{tbl:SR-examples}, $m$ denotes the number of rows in the parity-check matrix used).

We calculate the aforementioned bounds for the following codes:
\begin{itemize}
	\item the $[24,12,8]$ binary extended Golay code (cf. \cref{sec:Golay-numerics});
	\item the $[48,24,12]$ binary quadratic residue (QR) extended code (cf. \cite[Sec.~16]{macwilliams1977theory});
	\item the $(3,5)$-regular LDPC $[155,64,20]$ Tanner code in \cite{tanner01}.
\end{itemize}

\cref{tbl:SR-examples} presents numerical results. The original bound by Schwartz and Vardy \cref{eq:SV-bound} is the only constructive bound here, but it is the weakest one. Note that the bound in \cref{thm:sr-itw} is only slightly worse than \cite[Thm.~1]{yakimenka-skachekITW} but it is applicable to any code. Often, a code that does not satisfy \cref{eq:burdensome-req} has its stopping distance equal to the minimum distance. Yet the new bound is useful for calculation of the stopping redundancy hierarchy (see \cref{sec:SRH}).

The bounds in \cite[Thm.~1]{yakimenka-skachekITW} and \cref{thm:sr-itw} with $\tau=m$ give the tightest results. However, they require knowledge of the stopping sets spectrum of a parity-check matrix. For the Golay and the QR codes, we calculate their spectra by exhaustive brute-force checking. For the Tanner code, we use the spectrum obtained in \cite[Tab.~1]{rosnes2009efficient}. For longer codes, calculating a stopping sets spectrum can be infeasible even for the method in \cite{rosnes2009efficient} and similar works. We suggest a way to overcome this obstacle in \cref{sec:estimating-ui}.

\begin{table}
	\centering
	\caption{Comparison of upper bounds on stopping redundancy of different codes}
	\label{tbl:SR-examples}
	
	\renewcommand{\arraystretch}{1.5}
	\begin{tabular}{@{}lrrr@{}}
		\toprule
		& \begin{tabular}{@{}c@{}}$[24,12,8]$\\Golay\end{tabular} & \begin{tabular}{@{}c@{}}$[48,24,12]$\\QR\end{tabular} & \begin{tabular}{@{}c@{}}$[155,64,20]$\\Tanner\end{tabular} \\
		\midrule
		Equation \cref{eq:SV-bound}, \cite{schwartz-vardy2006} & \np{2509} & \np{4540385} & \np{6.2e18} \\
		Equation \cref{eq:HS-bound}, \cite{han2007improved} &  \np{232} & \np{4440} & \np{1526972} \\
		\cite[Thm.~7]{han2008improved} & \np{182} & \np{3564} & \np{1260673}\\
		\cite[Thm.~1]{yakimenka-skachekITW}, $\tau=1$ & 180 & 3538 & \np{1247888}\\
		\cref{thm:sr-itw}, $\tau=1$ & 185 & 3562 & \np{1247960} \\
		\cite[Thm.~1]{yakimenka-skachekITW}, $\tau=m$ & 168 & 2543 & \np{2573}\\
		\cref{thm:sr-itw}, $\tau=m$ & 168 & 2543 & \np{2573} \\
		\bottomrule
	\end{tabular}
	\renewcommand{\arraystretch}{1}
\end{table}

\subsection{Stopping Redundancy Hierarchy}\label{sec:SRH}
In \cref{def:stopping-redundancy}, it is required that the stopping distance of a parity-check matrix is exactly $d$. However, a more general requirement can be imposed. Thus, in \cite{olgica2008permutation}, it was required that the parity-check matrix does not contain stopping sets of size up to $\ell$, for some $\ell < d$. This can be achieved by adjoining a smaller number of rows to a parity-check matrix. 

It is important to notice that stopping sets of size $d$ or larger can also cause failures of the iterative decoder on the BEC (see, for example,~\cite{weber2005stopping}). Thus, in order to approach the ML performance with the iterative decoder, we should also cover the stopping sets of size $d$ or larger which, if erased, do not cause the ML decoder to fail. We can achieve this by adjoining redundant rows. The following definition is a generalization of \cite[Def.~2.4]{olgica2008permutation}. 

\begin{define}\label{def:stop-hier}
The $\ell$-th \emph{stopping redundancy} of $\code$, $1 \le \ell \le r$, $r = n-k$, is the smallest nonnegative integer $\rho_\ell(\code)$ such that there exists a (possibly redundant) parity-check matrix of $\code$ with $\rho_\ell(\code)$ rows and no stopping sets of size less than or equal to $\ell$ which are coverable by $\code$. The ordered set of integers $(\rho_1(\code), \rho_2(\code), \dotsc, \rho_{r}(\code))$ is called the \emph{stopping redundancy hierarchy} of $\code$.
\end{define}

The difference with \cite[Def.~2.4]{olgica2008permutation} is that, first, in \cref{def:stop-hier}, $\ell$ can be as large as $r$ (while in \cite[Def.~2.4]{olgica2008permutation}, $\ell \le d-1$, which is a more limiting condition). Second, in \cref{def:stop-hier}, only \emph{coverable} stopping sets are eliminated. However, as it was already mentioned, all stopping sets of size $\ell \le d-1$ are coverable, and therefore \cref{def:stop-hier} contains \cite[Def.~2.4]{olgica2008permutation} as a special case. Specifically, from \cref{def:stop-hier}, we have that $\rho(\code) = \rho_{d-1}(\code)$.

%\begin{note}
%	Our indexing is slightly different from that used in \cite{olgica2008permutation} and \cite{yakimenka-skachekITW}.
%\end{note}
\begin{note}
For $\ell \le d-1$, an upper bound on the $\ell$-th stopping redundancy can be formulated as in \cref{thm:sr-itw}, where $d$ is replaced by $\ell+1$. We omit the details.
\end{note}

\subsection{Choice of $H^{(\tau)}$}
\Cref{thm:sr-itw}  does not suggest how one should choose the initial $\tau \times n$ matrix. In general, it is a difficult question, as it strongly depends on the particular code. Below, we propose some simple heuristics.

Fix $\tau=1$. Then, \cref{lem:w-opt} in \cref{app:w-opt} gives two values for a weight $w$ of the initial row of $H^{(\tau)}$, one of which is guaranteed to cover the maximum number of stopping sets of size not more than $\ell$:
\[
w_{\mathrm{opt}} \in \left\{ \left\lfloor \frac{n+1}{\ell} \right\rfloor, \left\lceil \frac{n}{\ell} \right\rceil \right\} .
\]

However, a codeword of such weight does not necessarily exist in $\dual$. Hence one needs to consider the closest alternatives. After the dual codeword of weight $w$ is fixed, the number of stopping sets of size less than $d$ in $H^{(\tau)}$ is expressed as
\[
	u_i = \binom ni - w \binom{n-w}{i-1},
\]
and these values can be further used with the bound \cref{eq:ub-ITW}.

The situation becomes more complicated for $\tau = 2$, as in that case the optimal choice depends not only on the weights of the first two rows of $H^{(\tau)}$ but also on the size of the intersection of their supports. For simplicity, we can take two different rows of the same weight and obtain the corresponding estimate on the number of stopping sets. More precisely, if $\tau=2$, $H^{(\tau)}$ consists of two dual codewords $\vec h_1$ and $\vec h_2$ of weight $w$ each with an intersection of supports of size $|\supp \vec (h_1) \cap \supp \vec (h_2)| = \delta$, then the total number of stopping sets of size less than $d$ in $H^{(\tau)}$ equals (cf. \cite[Cor.~2]{yakimenka-skachekITW})
\ifonecolumn
\begin{align*}
u_i &= \binom ni - 2w \binom{n-w}{i-1} + \delta\binom{n-2w+\delta}{i-1} + (w-\delta)^2\binom{n-2w+\delta}{i-2}.
\end{align*}
\else
\begin{align*}
u_i = \binom ni &- 2w \binom{n-w}{i-1} + \delta\binom{n-2w+\delta}{i-1}\\
                &+ (w-\delta)^2\binom{n-2w+\delta}{i-2}.
\end{align*}
\fi

We can generalize this approach for $\tau > 2$ rows in $H^{(\tau)}$ by using the principle of inclusion-exclusion. However, this leads to explosion of terms in the formula for $u_i$. We do not continue in that direction.

%%%%%%%%%%%%%%%%%%%%%%%%%%%%%%%%%%%%%%%%%%%%%%%%%%%%%%%%%%%%%%%%%%%%%%%%%%%%%%%%%%%%%%%%%
\section{Achieving ML Performance}
\label{sec:ML-perf}

The ML decoder for the BEC is equivalent to solving a system of linear equations. More precisely, assume that we have a code with a parity-check matrix $H$, and that the received word is $\vec c$. Let the positions of erasures be $\cE \subseteq [n]$. Denote by $H_{\cE}$ the matrix formed by the columns of $H$ indexed by $\cE$, and by $\vec c_{\cE}$ denote the vector formed by the entries of $\vec c$ indexed by $\cE$. Denote $\bar \cE = [n] \setminus \cE$, and, similarly, define $H_{\bar \cE}$ and $\vec c_{\bar \cE}$. Then the parity-check equations can be written as
\[
H_{\cE} \vec c_{\cE} + H_{\bar \cE} \vec c_{\bar \cE} = \vec 0,
\]
where $\vec 0$ is the all-zero vector of the corresponding length. Since $\vec c_{\bar \cE}$, $H_{\bar \cE}$, and $H_{\cE}$ are known, we can rewrite the equations in the following form:
\begin{equation}
H_{\cE} \vec c_{\cE} = H_{\bar \cE} \vec c_{\bar \cE} \; . 
\label{eq:system_decode}
\end{equation}
It is a system of linear equations with the vector of unknowns $\vec c_{\cE}$ and a matrix of coefficients $H_{\cE}$. This system always has at least one solution, the originally transmitted codeword. If this solution is not unique, we say that the ML decoder fails.

The reason for the difference in performance of the ML and the iterative decoders is existence of stopping sets in a parity-check matrix used for iterative decoding. In the following sections, we aim at making the iterative decoding performance closer to that of the ML performance.

\subsection{Coverable Stopping Sets}
In \cref{sec:stop_redundancy}, we analyzed techniques for removal of all stopping sets of size up to $d-1$.
However, stopping sets of size $d$ or larger can also cause failures of the iterative decoder on the BEC. 
As it is mentioned above, in order to approach the ML performance with iterative decoding, one should also cover stopping sets of size $d$ or larger which, if erased, do not cause the ML decoder to fail. This can be achieved by adjoining redundant rows to a parity-check matrix. The following two lemmas will be instrumental in the analysis that follows.

As before, let $H$ be the parity-check matrix of the code $\code$. By $H^{(2^r)}$ we denote the matrix whose rows are all $2^r$ codewords of $\dual$, and $H^{(2^r)}_{\cE}$ denotes the matrix formed from columns of $H^{(2^r)}$ indexed by $\cE$. The next lemma consists of well-known results, see for example \cite[Sec.~3.2.1]{richarson2008}.

\begin{lemma}
	The following statements are equivalent:
	\begin{enumerate}[label=\roman*)]
		\item columns of $H_{\cE}$ are linearly dependent;
		\item there exists a non-zero codeword $\vec c$, such that $\operatorname{supp}(\vec c) \subseteq \cE$;
		\item if all positions in $\cE$ have been erased then the ML decoder fails.
	\end{enumerate}
\end{lemma}

Next, consider the case when the columns of $H_{\cE}$ are linearly independent.

\begin{lemma}\label{lem:recoverable-orthogonal}
	The following statements are equivalent:
	\begin{enumerate}[label=\roman*)]
		\item columns of $H_{\cE}$ are linearly independent;
		\item $H^{(2^r)}_{\cE}$ is an orthogonal array of strength $|\cE|$.
	\end{enumerate}
	And if any of them holds then
	\begin{enumerate}[resume,label=\roman*)]
		\item $\cE$ is not a stopping set in $H^{(2^r)}$.
	\end{enumerate}
\end{lemma}

\begin{IEEEproof}
	Both statements i) and iii) follow from ii) in a straightforward manner.  
	
	We prove next that ii) follows from i). First of all, if there are redundant rows in $H$, we can ignore them and assume that $m=r$. Owing to the fact that columns of $H_{\cE}$ are linearly independent, there exist $|\cE|$ rows in $H_{\cE}$ that form a full-rank square matrix. Then, each of the remaining $r - |\cE|$ rows of $H_{\cE}$ can be represented as a linear combination of these $|\cE|$ rows. Without loss of generality assume that
	\[
	H_{\cE} = \left(
	\begin{array}{c}
	B \\
	\hline
	TB
	\end{array}
	\right) ,
	\]
	where $B$ is an $|\cE| \times |\cE|$ full-rank matrix, and $T$ is a $(r - |\cE|) \times |\cE|$ matrix of coefficients.
	
	Each row of $H^{(2^r)}_{\cE}$ is bijectively mapped onto $r$ coefficients of linear combination $\vec \alpha = (\vec \alpha' \mid \vec \alpha'')$, where $\vec \alpha' \in \mathbb F_2^{|\cE|}$, and $\vec \alpha'' \in \mathbb F_2^{r - |\cE|}$, as follows:
	\[
	\vec \alpha \left(
	\begin{array}{c}
	B \\
	\hline	
	TB
	\end{array}
	\right)
	=
	\vec \alpha'B + \vec \alpha''TB = (\vec \alpha' + \vec \alpha''T)B .
	\]
	Fix the vector $\vec \alpha''$ (and therefore the vector $\vec \alpha''T$ of size $|\cE|$ is fixed). Then, the transformation
	\[
	\vec \alpha' \mapsto \vec \alpha' + \vec \alpha''T
	\]
	is a bijection of $\mathbb F_2^{|\cE|}$. Since $B$ is a full-rank matrix,
	\[
	\vec\alpha' \mapsto (\vec\alpha' + \vec\alpha''T)B
	\]
	is a bijection too. Hence, for a fixed $\vec\alpha''$, if we iterate over all $\vec\alpha'$, each of the rows in $\mathbb F_2^{|\cE|}$ is generated exactly once. This holds for each of $2^{r - |\cE|}$ possible choices for $\vec\alpha''$. Hence, each vector of $\mathbb F_2^{|\cE|}$ appears as a row in $H^{(2^r)}_{\cE}$ exactly $2^{r - |\cE|}$ times. Thus, $H^{(2^r)}_{\cE}$ is an orthogonal array of strength $|\cE|$.
\end{IEEEproof}
\medskip 

All things considered, an erasure pattern $\cE$ can be tackled by the ML decoder if and only if there exists a redundant row that covers $\cE$. Moreover, in this case $H^{(2^r)}_{\cE}$ is an orthogonal array of strength $|\cE|$ and, therefore, the techniques for calculating probability of being covered in the proof of \cref{thm:sr-itw} are still applicable. In the sequel, we reformulate the upper bound \cref{eq:ub-ITW}. 

%\begin{define}
%	We call a stopping set $\cS$ \emph{coverable} (by the code $\code$), if there is an (extended) parity-check matrix of $\code$ %in which $\cS$ is not a stopping set.
%\end{define}

Recall that the stopping set $\cS$ is coverable if there is a parity-check matrix in which $\cS$ is not a stopping set. As we see, coverable stopping sets are exactly those that, if erased, can be recovered by the ML decoder. 
%We can extend \cref{def:stop-hier} to stopping sets of size up to $r = n-k$.

%\begin{define}
%	For $\ell \leq r$, the $\ell$-th \emph{stopping redundancy} of $\code$ is the smallest nonnegative integer $\rho_\ell(\code)$ %such that there exists a (redundant) parity-check matrix of $\code$ with $\rho_\ell(\code)$ rows and no stopping sets of size %less than or equal to $\ell$ which are coverable by $\code$. The ordered set of integers $(\rho_1(\code), \rho_2(\code), \dotsc, %\rho_r(\code))$ is called the \emph{stopping redundancy hierarchy} of $\code$.
%\end{define}

We note that the $r$-th stopping redundancy $\rho_r(\code)$ of $\code$ is the smallest number of rows in a parity-check matrix of $\code$ such that the iterative decoder achieves the ML decoding performance. Next, we formulate an upper bound on the $\ell$-th stopping redundancy, as defined in \cref{def:stop-hier}, for $\ell \le r$, $r = n-k$.

\begin{thm}
	\label{thm:sr-ML}
	For an $[n,k,d]$ linear code $\code$ let $H^{(\tau)}$ be any $\tau \times n$ matrix consisting of $\tau$ different non-zero codewords of the dual code $\dual$ and let $u_i$ denote the number of not covered stopping sets of size $i$, $i=1,2,\dotsc,\ell$ ($\ell \leq r$), in $H^{(\tau)}$ that are coverable by $\code$. Then the $\ell$-th stopping redundancy is
	\begin{equation*}
	\rho_{\ell}(\code) \leq \Xi_{\ell}^{(I)} \left( u_1, u_2, \dotsc, u_{\ell} \right)
	\triangleq 
	\tau + \min_{0 \leq t < 2^r-\tau} \{t + \kappa_t\} + \Delta ,
	\end{equation*}
	where
	\begin{gather*}
		\mathcal D_t = \sum_{i = 1}^{\ell} u_i \prod_{j = \tau + 1}^{\tau + t} \pi(r,i,j) , \\
		P_{t,0} = \lfloor \mathcal D_t \rfloor , \\
		P_{t,j} = \Big\lfloor \pi(r, \ell, \tau+t+j) P_{t,j-1} \Big\rfloor , \quad j=1,2,\dotsc \\
		\Delta = r - \max \left\{ \rank H^{(\tau)}, \ell \right\} ,
	\end{gather*}
	and $\kappa_t$ is the smallest $j$ such that $P_{t,j} = 0$.
\end{thm}

We remark that the difference between the statements of \cref{thm:sr-ML} and of \cref{thm:sr-itw} is that the value $d-1$ is replaced by $\ell$.

\begin{IEEEproof}
	The proof follows the lines of that in \cref{thm:sr-itw} with the only difference that now for each coverable stopping set $\cS$, the corresponding matrix $H^{(2^r)}_\cS$ contains all the tuples of size $|\cS|$ equal number of times, as it was shown in \cref{lem:recoverable-orthogonal}.

We also observe that if, after adjoining rows, the matrix $H^{(\tau)}$ does not contain coverable stopping sets of size at most $\ell$, $\ell \le r$, then the rank of this matrix is at least $\ell$. This follows from the fact that there exist $\ell$ 
linearly independent columns in the resulting matrix, as otherwise the corresponding set of coordinates would contain a stopping set that is not a support of a codeword.

\end{IEEEproof}

\begin{cor}
	There exists an extended parity-check matrix with no more than $\Xi_{r}^{(I)} \left( u_1, u_2, \dotsc, u_{r} \right)$ rows, such that the iterative decoder fails if and only if the ML decoder fails. It follows that the decoding error probability of these two decoders is equal.
\end{cor}

Computing the number $u_i$ of stopping sets of size $i$---or even finding the corresponding upper bound---might be a difficult task for general codes, except for trivial cases. In what follows, we suggest two approaches:
\begin{itemize}
	\item ensemble-average approach (see \cref{sec:ens-ave-ML-SR});
	\item finding estimates on $u_i$ numerically (see \cref{sec:estimating-ui}).
\end{itemize}

\subsection{Exact Ensemble-Average ML Stopping Redundancy}\label{sec:ens-ave-ML-SR}
In order to apply the upper bounds on the stopping redundancy to the ensemble-average values, we formulate a weaker bound inspired by \cite{han2008improved}.

\begin{thm}\label{thm:sr-ML-better-convex}
	Assume that $\code$ is a linear $[n,k]$-code and $H$ is a parity-check matrix consisting of $m$ different rows being codewords of the dual code $\dual$, such that there are $u_i$ stopping sets of size $i=1, 2, \dotsc, \ell$ ($\ell \leq r$), in $H$ coverable by $\code$. Then the $\ell$-th stopping redundancy is bounded from above as follows:
	\ifonecolumn
	\begin{equation*}
	\rho_{\ell} (\code) \leq \Xi_{\ell}^{(II)} \left(u_1, u_2, \dotsc, u_{\ell} \right) \\
	\triangleq
	m + \min_{0 \leq t < 2^m-m} \left\{ t + \sum_{i=1}^{\ell} u_i \prod_{j=m+1}^{m+t} \pi(m,i,j) \right\}.
	\end{equation*}
	\else
	\begin{multline*}
	\rho_{\ell} (\code) \leq \Xi_{\ell}^{(II)} \left(u_1, u_2, \dotsc, u_{\ell} \right) \\
	\triangleq
	m + \min_{0 \leq t < 2^m-m} \left\{ t + \sum_{i=1}^{\ell} u_i \prod_{j=m+1}^{m+t} \pi(m,i,j) \right\}.
	\end{multline*}
	\fi
\end{thm}
\begin{IEEEproof}
	Analogous to Step 1 in \cref{thm:sr-itw}, choose $t$, $t=0,1,\dotsc,2^r-m-1$, codewords from $\dual_0 \setminus \{\text{rows of } H\}$ uniformly at random and without repetitions, and adjoin them to $H$. Then, the average number of coverable but not covered stopping sets in this extended matrix becomes equal to
	\[
	\sum_{i=1}^{\ell} u_i \prod_{j=m+1}^{m+t} \pi(r,i,j).
	\]
	
	For each of these stopping sets, we add one row from $\dual_0$ to cover it, and thus the total number of rows in the parity-check matrix becomes
	\ifonecolumn
	\begin{equation*}
		m + t + \sum_{i=1}^{\ell} u_i \prod_{j=m+1}^{m+t} \pi(r,i,j) \leq m + t + \sum_{i=1}^{\ell} u_i \prod_{j=m+1}^{m+t} \pi(m,i,j) .
	\end{equation*}
	\else
	\begin{multline*}
	m + t + \sum_{i=1}^{\ell} u_i \prod_{j=m+1}^{m+t} \pi(r,i,j) \\
	\leq m + t + \sum_{i=1}^{\ell} u_i \prod_{j=m+1}^{m+t} \pi(m,i,j) .
	\end{multline*}
	\fi
By minimizing this expression over the choice of $t$, we obtain the required upper bound. We note that minimizing over $t$ up to $2^m-m$ is just a matter of further convenience, as the true minimum value is obtained for $t < 2^r-m \leq 2^m-m$.
\end{IEEEproof}

Now we can formulate the ensemble-average result.

\begin{cor}\label{cor:rho-ens-average}
Consider an ensemble $\ensemble$ of codes, where the probability distribution of the codes is determined by the probability distribution on $m \times n$ parity-check matrices. Moreover, assume that the parity-check matrix $H$  of rank $r = n - k$ corresponding to the $[n,k]$ code $\code \in \ensemble$ has $u^{(H)}_i$ size-$i$ stopping sets coverable by $\code$, where $i = 1, 2, \cdots, \ell$. Denote the ensemble-average number of such stopping sets:
	\[
	\bar u_i = \Expect[\ensemble]{u_i^{(H)}} .
	\]
	
	Then, the average $\ell$-th stopping redundancy over the ensemble $\ensemble$ is bounded from above as follows:
	\begin{equation*}
		\Expect[\ensemble]{ \rho_{\ell}(\code) } \leq 
		\Xi_{\ell}^{(II)} (\bar u_1, \bar u_2, \dotsc, \bar u_{\ell}).
	\end{equation*}
\end{cor}
\begin{IEEEproof}
	First, we observe that \cref{thm:sr-ML-better-convex} yields an upper bound on $\rho_{\ell} (\code)$ for every integer $0 \leq t < 2^m-m$:
	\[
	\rho_{\ell} (\code) \leq m + t + \sum_{i=1}^{\ell} u_i \prod_{j=m+1}^{m+t}\pi(m,i,j).
	\] 
	Then, $\Xi_{\ell}^{(II)}$ is a minimum of these upper bounds over the values of $t$.
	
	Fix some integer $0 \leq t < 2^m - m$ and take the average over $\ensemble$:
	\begin{equation*}
	\Expect[\ensemble]{ \rho_{\ell} (\code) } 
	\leq m + t + \sum_{i=1}^{\ell} \bar u_i \prod_{j=m+1}^{m+t} \pi(m,i,j).
	\end{equation*}

	As it holds for each $t$, it should also hold for their minimum:
	\begin{equation*}
	\Expect[\ensemble]{ \rho_{\ell} (\code) } \leq 
	m + \min_{0 \leq t < 2^m-m} \left\{ t + \sum_{i=1}^{\ell} \bar u_i \prod_{j=m+1}^{m+t} \pi(m,i,j) \right\}.
	\end{equation*}
\end{IEEEproof}

\subsection{Statistical Estimation of the Number of Coverable Stopping Sets}\label{sec:estimating-ui}
In this section, we aim at finding statistical estimates on the number of coverable stopping sets and further apply them to the upper bounds on the stopping redundancy hierarchy. In what follows, we use the cumulative distribution function of the standard normal distribution $\mathcal N(0, 1)$ given by:
\[
\Phi(x) = \frac{1}{\sqrt{2 \pi}} \int_{-\infty}^x e^{-t^2/2} \mathrm d t .
\]

\begin{lemma}\label{lem:conf-interval-H}
Consider a parity-check matrix $H$ of an $[n,k]$-code $\code$. For $1 \leq i \leq r$, fix a large number $N_i$ and generate $N_i$ random subsets of $[n]$ uniformly at random (with repetitions), namely $\cS_1^{(i)}, \cS_2^{(i)}, \dotsc, \cS_{N_i}^{(i)}$, each subset consisting of $i$ elements. For  $j = 1, 2, \dotsc, N_i$, we define the following events:
\[
x_j^{(i)} = \begin{cases}
1, & \text{if $\cS_j^{(i)}$ is a coverable stopping set in $H$,} \\
0, & \text{otherwise.} 
\end{cases}
\]
If $u_i$ is a number of size-$i$ stopping sets in $H$ coverable by $\code$, and $\epsilon_i$ is some small fixed number, then%\footnote{{We note that in fact this probability is approximate but it becomes exact for $N_i \rightarrow \infty$. We refer interested reader to \cite{cai2005one} and the references therein for more details.}}
\[
\Prob[]{u_i < \hat u_i} = 1-\epsilon_i,
\]
where
\begin{gather}
\hat u_i = \binom ni \left( \tilde x^{(i)} + \kappa \sqrt{\frac{\hat V}{N_i} + \frac{\gamma_1 \hat V + \gamma_2}{N_i^2}} \right),\label{eq:eq_1} \\
\kappa = \Phi^{-1}(1-\epsilon_i), \qquad \eta = \nicefrac{\kappa^2}{3} + \nicefrac{1}{6}, \\
\bar x^{(i)} = \frac{\sum_{j=1}^{N_i}x_j^{(i)}}{N_i}, \qquad 
\tilde x^{(i)} = \frac{N_i \bar x^{(i)} + \eta}{N_i + 2\eta}, \\
\gamma_1 = \nicefrac{-13}{18} \cdot \kappa^2 - \nicefrac{17}{18}, \qquad \gamma_2 = \nicefrac{\kappa^2}{18} + \nicefrac{7}{36}, \\
\hat V = \bar x^{(i)}(1-\bar x^{(i)})\label{eq:eq_last} \; .
\end{gather}
\end{lemma}

\begin{IEEEproof}
	Random variables $\{x_j^{(i)}\}$ are independent and identically distributed according to the Bernoulli distribution with success probability
	\[
	\theta_i = \frac{u_i}{\binom ni} .
	\]
	Here $\theta_i$ is unknown because $u_i$ is unknown.
	
	We further apply \emph{the $1-\epsilon_i$ upper limit second-order corrected one-sided confidence interval}, constructed in \cite[(10)]{cai2005one} and based on Edgeworth expansion. In our notation, it states that
	\begin{equation}\label{eq:conf-interval}
	\Prob{\theta_i < \tilde x^{(i)} + \kappa \sqrt{\frac{\hat V}{N_i} + \frac{\gamma_1 \hat V + \gamma_2}{N_i^2}}} = 1-\epsilon_i.
	\end{equation}
	
	From this we obtain the required result.
\end{IEEEproof}

This estimate can be used in conjunction with the upper bounds in \cref{thm:sr-ML} and \cref{thm:sr-ML-better-convex}. More specifically, we fix $N_1, N_2, \dotsc, N_\ell$, and $\epsilon_1, \epsilon_2, \dotsc, \epsilon_{\ell}$, and then we obtain that
\[
\rho_{\ell}(\code) \leq \Xi_{\ell}^{(I)} \left( \hat u_1, \hat u_2, \dotsc, \hat u_{\ell} \right),
\]
which holds with probability
\[
\prod_{i=1}^{\ell} (1-\epsilon_i) .
\]

Furthermore, this approach can be extended to estimating the ensemble-average $\ell$-th stopping redundancy, $\Expect[\ensemble]{ \rho_{\ell}(\code) }$.

\begin{lemma}\label{lem:conf-interval-ens}
	In the settings of \cref{cor:rho-ens-average}, for $1 \leq i \leq m$, fix a large number $N_i$ and generate $N_i$ random pairs $\left(H_j^{(i)}, \cS_j^{(i)}\right)$, $j=1,2,\dotsc,N_i$, where $H_j^{(i)}$ is a parity-check matrix of a code from $\ensemble$ and $S_j^{(i)}$ is a random subset of $[n]$ consisting of $i$ elements, $H_j^{(i)}$ and $\cS_j^{(i)}$ being independent.
	
	For  $j = 1, 2, \dotsc, N_i$, we define the following events:
	\[
	y_j^{(i)} = \begin{cases}
	1, & \text{if $\cS_j^{(i)}$ is a coverable stopping set in $H_j^{(i)}$,} \\
	0, & \text{otherwise.} 
	\end{cases}
	\]
	
	For a fixed small $\epsilon_i$,
	\[
	\Prob[]{\bar u_i < \hat{\bar u}_i} = 1 - \epsilon_i,
	\]
	where $\hat{\bar u}_i$ is defined similar to $\hat{u}_i$ in \crefrange{eq:eq_1}{eq:eq_last} with
	$x_j^{(i)}$, $\bar x^{(i)}$ and $\tilde x^{(i)}$ replaced by $y_j^{(i)}$, $\bar y^{(i)}$ and $\tilde y^{(i)}$, respectively.  

%	\begin{gather*}
%	\hat{\bar u}_i = \binom ni \left( \tilde y^{(i)} + \kappa \sqrt{\frac{\hat V}{N_i} + \frac{\gamma_1 \hat V + \gamma_2}{N_i^2}} %\right), \\
%	\kappa = \Phi^{-1}(1-\epsilon_i), \qquad \eta = \kappa^2 / 3 + 1/6, \\
%	\bar y^{(i)} = \frac{\sum_{j=1}^{N_i}y_j^{(i)}}{N_i}, \qquad 
%	\tilde y^{(i)} = \frac{N_i \bar y^{(i)} + \eta}{N_i + 2\eta}, \\
%	\gamma_1 = -\frac{13}{18} \kappa^2 - \frac{17}{18}, \qquad \gamma_2 = \frac{\kappa^2}{18} + \frac{7}{36}, \\
%	\hat V = \bar y^{(i)}(1-\bar y^{(i)}).
%	\end{gather*}
\end{lemma}

\begin{IEEEproof}
	Analogous to the proof of \cref{lem:conf-interval-H}.
\end{IEEEproof}
\medskip

If we fix $N_1, N_2, \dotsc, N_{\ell}$ and $\epsilon_1, \epsilon_2, \dotsc, \epsilon_{\ell}$, we obtain that
\begin{equation}\label{eq:expected_rho}
	\Expect[\ensemble]{ \rho_{\ell}(\code) } \leq 
	\Xi_{\ell}^{(II)} (\hat{\bar u}_1, \hat{\bar u}_2, \dotsc, \hat{\bar u}_{\ell})
\end{equation}
with probability $\prod_{i=1}^{\ell}(1-\epsilon_i)$.

\subsection{Case Study: Standard Random Ensemble}
\label{sec:SRE-casestudy}
In this section, we demonstrate application of the aforementioned bounds to the standard random ensemble (SRE) $\mathfrak{S}(n, m)$. This ensemble is defined by a means of its $m \times n$ parity-check matrices $H$, where each entry of $H$ is an independent and identically distributed (i.i.d.) Bernoulli random variable with parameter $1/2$.

As it is shown in \cref{app:alekseyev}, for $i \leq m$, the number of full-rank $m \times i$ matrices with no rows of weight one is equal to $\mathcal N(m, i)$ defined in \cref{eq:N}. Fix some subset of columns $\cS$ of size $i$, and choose a random parity-check matrix $H$ from the ensemble $\mathfrak S(n, m)$. The probability that there is a coverable (but not covered) stopping set in the columns indicated by $\cS$, is as follows:
\begin{equation}
	\frac{\mathcal N(m, i)}{2^{mi}} .
\label{eq:fraction}
\end{equation}
We used here the fact that $H_\cS$, the submatrix of $H$ consisting of columns indexed by $\cS$, is equal to every $m \times i$ matrix equiprobably. Therefore, the average number of coverable but not covered stopping sets of size $i$ in $H$ is
\[
	\bar u_i = \Expect[\mathfrak S(n, m)]{u_i^{(H)}}  = \binom ni \frac{\mathcal N(m, i)}{2^{mi}} .
\]

Next, we can apply \cref{cor:rho-ens-average} to obtain the upper bound on the ensemble-average $\ell$-th stopping redundancy.

\begin{figure}[bt!]
\begin{center}
	\includegraphics[scale=.9]{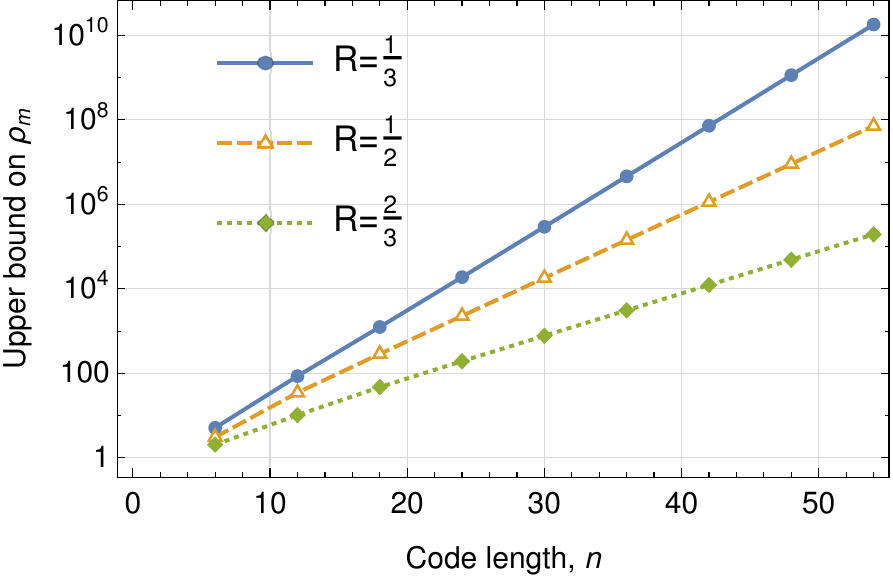}
	\caption{Upper bounds on $\mathfrak{S}(n,m)$-average $m$-th stopping redundancy ($m = (1 - R)n$).}
	\label{fig:SRE-plot}
\end{center}
\end{figure}

We illustrate the behavior of the obtained bound in \cref{fig:SRE-plot}. It can be observed empirically that the bound grows exponentially. We remark that the presented values of the upper bound on the maximal stopping redundancy 
(\cref{fig:SRE-plot}, \cref{tbl:SRE-simulations}, \cref{fig:GAL-plot})
in some cases can take on very large values. In this work, we only show consistency of the obtained numerical results and the theoretical bounds. However, our experiments with short to moderate length codes~\cite{vihula, isit2018} show that decoding with redundant parity-check matrices can be a practical near-ML decoding technique in some cases.  

%\subsection{Case Study: Gallager Ensemble}
%Here we consider an average stopping redundancy over the Gallager ensemble $\mathfrak{Gal}(n, J, K)$ of $(J,K)$-regular LDPC codes \cite{gallager1962low}. The ensemble is defined by parity-check matrices of special form. An $\left( \frac{n J}K \right) \times n$ parity-check matrix consists of $J$ strips of width $M = n/K$ rows each. In the first strip, the $j$th row contain $K$ ones in positions $(j-1)K+1, (j-1)K+2, \dotsc, jK$ for $j=1,2,\dotsc,M$. And each of the other strips is a random column permutation of the first strip.
%
%We were unable to find exact values for ensemble-average numbers of coverable stopping sets. Therefore, as an upper bound we will use ensemble-average spectrum of stopping sets (see e.g., \cite{averagespectra2017}). More precisely, we take
%\[
%\Expect[\mathfrak{Gal}(n,J,K)]{u^{(\code)}} = \sum_{i=1}^{\ell} \frac{\coef \left( \left( (1+s)^K - K s \right)^M, \, s^i \right)^J}{\binom ni^{J-1}} ,
%\]
%and further apply \cref{cor:rho-ens-average}.

%%%%%%%%%%%%%%%%%%%%%%%%%%%%%%%%%%%%%%%%%%%%%%%%%%%%%%%%%%%%%%%%%%%%%%%%%%%%%%%%%%%%%%%%%

\section{Numerical Results}
\label{sec:simulations}
\subsection{$[24,12,8]$ extended Golay code}\label{sec:Golay-numerics}
Consider the $[24,12,8]$ extended Golay code. We use the systematic double-circulant matrix $H$ given in \cite[p.~65]{macwilliams1977theory} as a means to define the code (see \cref{tbl:Golay-matrix}). The matrix has the stopping distance $s(H) = 4$.
	
\begin{table*}[!t]
	\caption{Parity-check matrix of the extended $[24,12,8]$ Golay code. Dots denote zeros.}
	\label{tbl:Golay-matrix}
	\[
	\begin{pmatrix}
	1 & 1 & \cdot & \cdot & \cdot & \cdot & \cdot & \cdot & \cdot & \cdot & \cdot & \cdot & \cdot & 1 & 1 & \cdot & 1 & 1 & 1 & \cdot & \cdot & \cdot & 1 & \cdot \\
	1 & \cdot & 1 & \cdot & \cdot & \cdot & \cdot & \cdot & \cdot & \cdot & \cdot & \cdot & \cdot & \cdot & 1 & 1 & \cdot & 1 & 1 & 1 & \cdot & \cdot & \cdot & 1 \\
	1 & \cdot & \cdot & 1 & \cdot & \cdot & \cdot & \cdot & \cdot & \cdot & \cdot & \cdot & \cdot & 1 & \cdot & 1 & 1 & \cdot & 1 & 1 & 1 & \cdot & \cdot & \cdot \\
	1 & \cdot & \cdot & \cdot & 1 & \cdot & \cdot & \cdot & \cdot & \cdot & \cdot & \cdot & \cdot & \cdot & 1 & \cdot & 1 & 1 & \cdot & 1 & 1 & 1 & \cdot & \cdot \\
	1 & \cdot & \cdot & \cdot & \cdot & 1 & \cdot & \cdot & \cdot & \cdot & \cdot & \cdot & \cdot & \cdot & \cdot & 1 & \cdot & 1 & 1 & \cdot & 1 & 1 & 1 & \cdot \\
	1 & \cdot & \cdot & \cdot & \cdot & \cdot & 1 & \cdot & \cdot & \cdot & \cdot & \cdot & \cdot & \cdot & \cdot & \cdot & 1 & \cdot & 1 & 1 & \cdot & 1 & 1 & 1 \\
	1 & \cdot & \cdot & \cdot & \cdot & \cdot & \cdot & 1 & \cdot & \cdot & \cdot & \cdot & \cdot & 1 & \cdot & \cdot & \cdot & 1 & \cdot & 1 & 1 & \cdot & 1 & 1 \\
	1 & \cdot & \cdot & \cdot & \cdot & \cdot & \cdot & \cdot & 1 & \cdot & \cdot & \cdot & \cdot & 1 & 1 & \cdot & \cdot & \cdot & 1 & \cdot & 1 & 1 & \cdot & 1 \\
	1 & \cdot & \cdot & \cdot & \cdot & \cdot & \cdot & \cdot & \cdot & 1 & \cdot & \cdot & \cdot & 1 & 1 & 1 & \cdot & \cdot & \cdot & 1 & \cdot & 1 & 1 & \cdot \\
	1 & \cdot & \cdot & \cdot & \cdot & \cdot & \cdot & \cdot & \cdot & \cdot & 1 & \cdot & \cdot & \cdot & 1 & 1 & 1 & \cdot & \cdot & \cdot & 1 & \cdot & 1 & 1 \\
	1 & \cdot & \cdot & \cdot & \cdot & \cdot & \cdot & \cdot & \cdot & \cdot & \cdot & 1 & \cdot & 1 & \cdot & 1 & 1 & 1 & \cdot & \cdot & \cdot & 1 & \cdot & 1 \\
	\cdot & \cdot & \cdot & \cdot & \cdot & \cdot & \cdot & \cdot & \cdot & \cdot & \cdot & \cdot & 1 & 1 & 1 & 1 & 1 & 1 & 1 & 1 & 1 & 1 & 1 & 1 \\
	\end{pmatrix}
	\]
\end{table*}

Due to the short code length, we are able to calculate the values ${u_1,u_2,\dotsc,u_{12}}$ by exhaustive checking of all the subsets of $\{1,2,\dotsc,24\}$ of size up to 12. We use these values to calculate the upper bounds in \cref{thm:sr-ML,thm:sr-ML-better-convex}.

Next, we generate $N_i = 1000$ ($1 \leq i \leq 12$) random subsets of $\{1, 2, \cdots, 24\}$ and register the events according to \cref{lem:conf-interval-H}. The following sequence of frequencies of coverable stopping sets (as defined in~\cref{lem:conf-interval-H}) was obtained:
\ifonecolumn
\begin{align*}
\left\{ \bar x^{(i)} \right\}_{i=1}^{12} = \{0, 0, 0, 0.01, 0.039, 0.122, 0.219, 0.345, 0.487, 0.621, 0.652, 0.463\} .
\end{align*}
\else
\begin{align*}
\left\{ \bar x^{(i)} \right\}_{i=1}^{12} = \{&0, 0, 0, 0.01, 0.039, 0.122, 0.219,\\
 &0.345, 0.487, 0.621, 0.652, 0.463\} .
\end{align*}
\fi

We repeat the experiments with a different value $N_i = 10^6$ ($1 \leq i \leq 12$), and obtain the following sequence of frequencies:
\ifonecolumn
\begin{align*}
\left\{ \bar x^{(i)} \right\}_{i=1}^{12} = \{0, 0, 0, 0.010314, 0.042985, 0.109956, 
0.214436, 0.350958, 0.496478, 0.616122, 0.635654, 0.440123\} .
\end{align*}
\else
\begin{align*}
	\left\{ \bar x^{(i)} \right\}_{i=1}^{12} = \{&0, 0, 0, 0.010314, 0.042985,
	0.109956,\\
	&0.214436, 0.350958, 0.496478, 0.616122,\\
	&0.635654, 0.440123\}.
\end{align*}

\fi

By setting $\epsilon_i = 0.001$ for all $i$ (therefore, $\prod_{i=1}^{12}(1-\epsilon_i) = 0.988066$), we employ both sets of values in \cref{lem:conf-interval-H} and, further, in \cref{thm:sr-ML,thm:sr-ML-better-convex}. The results are presented in \cref{tbl:golay-SR}. We observe consistency between the theoretical and the empirical results presented therein.

\begin{table}[!t]
	\centering
	\caption{Stopping redundancy hierarchies. The $[24,12,8]$ extended Golay code}
	\label{tbl:golay-SR}
	
	\renewcommand{\arraystretch}{1.2}
	\begin{tabular}{@{}lrrcrrcrr@{}}
		\toprule
		& \multicolumn{2}{c}{\begin{tabular}{@{}c@{}}coverable\\stopping sets\end{tabular}} & \phantom{.} & \multicolumn{2}{c}{\cref{thm:sr-ML}} & \phantom{.} & \multicolumn{2}{c}{\cref{thm:sr-ML-better-convex} }\\
		\cmidrule{2-3} \cmidrule{5-6} \cmidrule{8-9}
		%				{} & $i$ & $u_i$ && $\ell$ & $\rho_{\ell}$ && $\ell$ & $\rho_{\ell}$ \\
		%		\midrule
		\multirow{12}{*}{Exact $u_i$} & $u_1$ & 0 && $\rho_1$ & 12  && $\rho_1$ & 12 \\
		& $u_2$ & 0 && $\rho_2$ & 12  && $\rho_2$ & 12 \\
		& $u_3$ & 0 && $\rho_3$ & 12  && $\rho_3$ & 12 \\
		& $u_4$ & 110 && $\rho_4$ & 25  && $\rho_4$ & 27 \\
		& $u_5$ & \np{1837} && $\rho_5$ & 49  && $\rho_5$ & 51 \\
		& $u_6$ & \np{14795} && $\rho_6$ & 91  && $\rho_6$ & 95 \\
		& $u_7$ & \np{74349} && $\rho_7$ & 168  && $\rho_7$ & 174 \\
		& $u_8$ & \np{257796} && $\rho_8$ & 304  && $\rho_8$ & 316 \\
		& $u_9$ & \np{649275} && $\rho_9$ & 540  && $\rho_9$ & 560 \\
		& $u_{10}$ & \np{1206755} && $\rho_{10}$ & 927  && $\rho_{10}$ & 960 \\
		& $u_{11}$ & \np{1585794} && $\rho_{11}$ & \np{1507}  && $\rho_{11}$ & \np{1558} \\
		& $u_{12}$ & \np{1189574} && $\rho_{12}$ & \np{2241}  && $\rho_{12}$ & \np{2309} \\
		\midrule
		\multirow{12}{*}{\begin{tabular}{@{}c@{}}Estimates $\hat u_i$\\$(N_i=10^3)$\end{tabular}} & $\hat u_1$ & 0 && $\rho_1$ & 12 && $\rho_1$ & 12 \\
		& $\hat u_2$ & 1 && $\rho_2$ & 13 && $\rho_2$ & 13 \\
		& $\hat u_3$ & 12 && $\rho_3$ & 17 && $\rho_3$ & 17 \\
		& $\hat u_4$ & 247 && $\rho_4$ & 28 && $\rho_4$ & 30 \\
		& $\hat u_5$ & \np{2596} && $\rho_5$ & 51 && $\rho_5$ & 53 \\
		& $\hat u_6$ & \np{21061} && $\rho_6$ & 94 && $\rho_6$ & 98 \\
		& $\hat u_7$ & \np{90406} && $\rho_7$ & 171 && $\rho_7$ & 178 \\
		& $\hat u_8$ & \np{288582} && $\rho_8$ & 307 && $\rho_8$ & 319 \\
		& $\hat u_9$ & \np{700573} && $\rho_9$ & 544 && $\rho_9$ & 564 \\
		& $\hat u_{10}$ & \np{1309119} && $\rho_{10}$ & 933 && $\rho_{10}$ & 967 \\
		& $\hat u_{11}$ & \np{1740882} && $\rho_{11}$ & \np{1519} && $\rho_{11}$ & \np{1570} \\
		& $\hat u_{12}$ & \np{1384130} && $\rho_{12}$ & \np{2265} && $\rho_{12}$ & \np{2333} \\
		\midrule
		\multirow{12}{*}{\begin{tabular}{@{}c@{}}Estimates $\hat u_i$\\$(N_i=10^6)$\end{tabular}} & $\hat u_1$ & 0 && $\rho_1$ & 12 && $\rho_1$ & 12 \\
		& $\hat u_2$ & 0 && $\rho_2$ & 12 && $\rho_2$ & 12 \\
		& $\hat u_3$ & 0 && $\rho_3$ & 12 && $\rho_3$ & 12 \\
		& $\hat u_4$ & 112 && $\rho_4$ & 25 && $\rho_4$ & 27 \\
		& $\hat u_5$ & \np{1853} && $\rho_5$ & 49 && $\rho_5$ & 51 \\
		& $\hat u_6$ & \np{14930} && $\rho_6$ & 91 && $\rho_6$ & 95 \\
		& $\hat u_7$ & \np{74656} && $\rho_7$ & 168 && $\rho_7$ & 174 \\
		& $\hat u_8$ & \np{259204} && $\rho_8$ & 304 && $\rho_8$ & 316 \\
		& $\hat u_9$ & \np{651167} && $\rho_9$ & 540 && $\rho_9$ & 561 \\
		& $\hat u_{10}$ & \np{1211318} && $\rho_{10}$ & 927 && $\rho_{10}$ & 961 \\
		& $\hat u_{11}$ & \np{1590393} && $\rho_{11}$ & \np{1508} && $\rho_{11}$ & \np{1559} \\
		& $\hat u_{12}$ & \np{1194310} && $\rho_{12}$ & \np{2241} && $\rho_{12}$ & \np{2310} \\
		\bottomrule
	\end{tabular}
	\renewcommand{\arraystretch}{1}
\end{table}

\subsection{Greedy heuristics for Golay redundant parity-check matrices}
In \cite{schwartz-vardy2006}, the authors suggest a greedy (lexicographic) algorithm to search for redundant rows in order to remove all stopping sets of size up to $7$. The algorithm requires the full list of stopping sets, as well as the full list of dual codewords. We note that this straightforward approach is applicable to the aforementioned Golay code due to its short length.

Based on the ideas discussed in \cref{sec:ML-perf}, we can apply the algorithm akin to that of Schwartz and Vardy beyond the code minimum distance. In that case, the algorithm works with the full list of \emph{coverable} stopping sets of the code. We now describe the algorithm in more details. We use the systematic double-circulant matrix $H$ given in \cite[p.~65]{macwilliams1977theory} as a  means to define the Golay code. Its stopping distance $s(H) = 4$.

Fix $\ell$, $4 \leq \ell \leq 12$, and generate the list
\[
\mathcal L = \{ \mathcal S \subseteq [n] : |\mathcal S| \leq \ell, \, \rank H_{\mathcal S} = |\mathcal S| \} ,
\]
i.e., the list of stopping sets of size up to $\ell$ (incl.) coverable by the Golay code. Next, we iteratively construct the extended parity-check matrix, starting with the empty matrix. At each iteration, we find one of the 4095 non-zero dual codewords\footnote{Although in fact the $[24,12,8]$ Golay code is self-dual.} with the highest score. The score is of heuristic nature and for a dual codeword $\vec h$ it is calculated as follows:
\[
\operatorname{score}(\vec h) = \sum_{\mathcal S \in \mathcal L} |\mathcal S| \cdot \mathbb I \{\vec h \text{ covers } \mathcal S \},
\]
where $\mathbb I \{ \cdot \}$ is the indicator function. The row $\vec h^*$ with the maximum score is added to the matrix we build, and the stopping sets covered by $\vec h^*$ are removed from $\mathcal L$. Iterations continue until $\mathcal L$ is empty. As we have only coverable sets in $\mathcal L$, the algorithm will stop before we add all the 4095 rows. To this end, we verify that the obtained parity-check matrix has rank 12.

\begin{figure}
	\centering
	\includegraphics[scale=.9]{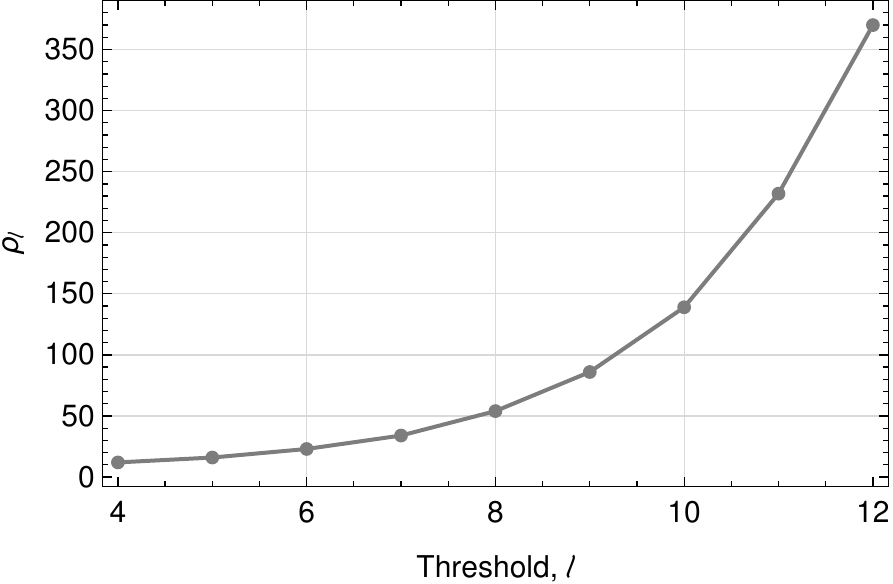}
	\caption{Estimate on stopping redundancy hierarchy obtained by greedy search}
	\label{fig:Golay-SV-SRH}
\end{figure}

A small difference with \cite{schwartz-vardy2006} in the proposed approach is a random choice of $\vec h^*$ when several dual codewords have the same score. In that case, we run the algorithm several times and choose the matrix with the least number of rows. \cref{fig:Golay-SV-SRH} illustrates the number of rows in the best obtained matrices for $\ell = 4,5,\dotsc,12$. We further refer to these matrices as $H^{(12)}$, $H^{(16)}$, $H^{(23)}$, $H^{(34)}$, $H^{(54)}$, $H^{(86)}$, $H^{(139)}$, $H^{(232)}$, and $H^{(370)}$, according to the number of rows they have. The notation $\Psi_{H}$ is used to denote the number of undecodable erasure patterns in the parity-check matrix $H$. 

\cref{tbl:Golay-red-patterns} shows the numbers of undecodable patterns for the aforementioned extended parity-check matrices. Note that the number of such patterns for the BP decoder with $H^{(370)}$ is \emph{exactly} the same as for the ML decoder. This is in accordance with the discussion in \cref{sec:ML-perf}.

\begin{table*}[!t]
	\centering
	\caption{Number of undecodable erasure patterns for different parity-check matrices of the $[24,12,8]$ Golay code}
	\label{tbl:Golay-red-patterns}
	
	\renewcommand{\arraystretch}{1.5}% Wider
	\begin{tabular}{@{}cccccccccccc@{}}
		\toprule
		\multirow{2}{*}{~} & \multicolumn{11}{c}{$w$} \\
		\cmidrule{2-12}
		& 0--3 & 4 & 5 & 6 & 7 & 8 & 9 & 10 & 11 & 12 & $\geq 13$ \\
		\midrule
		Total patterns &  & 10\,626 & 42\,504 & 134\,596 & 346\,104 & 735\,471 & 1\,307\,504 & 1\,961\,256 & 2\,496\,144 & 2\,704\,156 & $\binom{24}{w}$ \\
		$\Psi_{H}$ & 0 & 110 & 2\,277 & 19\,723 & 100\,397 & 343\,035 & 844\,459 & 1\,568\,875 & 2\,274\,130 & 2\,637\,506 & $\binom{23}{w}$ \\
		$\Psi_{H^{(34)}}$ & 0 & 0 & 0 & 0 & 0 & 3\,598 & 82\,138 & 585\,157 & 1\,717\,082 & 2556402 & $\binom{24}{w}$ \\
		$\Psi_{H^{(54)}}$ & 0 & 0 & 0 & 0 & 0 & 759 & 16\,424 & 195\,190 & 1\,027\,002 & 2\,242\,956 & $\binom{24}{w}$ \\
		$\Psi_{H^{(86)}}$ & 0 & 0 & 0 & 0 & 0 & 759 & 12\,144 & 98\,822 & 570\,567 & 1\,774\,724 & $\binom{24}{w}$ \\
		$\Psi_{H^{(139)}}$ & 0 & 0 & 0 & 0 & 0 & 759 & 12\,144 & 91\,080 & 437\,744 & 1\,438\,874 & $\binom{24}{w}$ \\
		$\Psi_{H^{(232)}}$ & 0 & 0 & 0 & 0 & 0 & 759 & 12\,144 & 91\,080 & 425\,040 & 1\,324\,074 & $\binom{24}{w}$ \\
		$\Psi_{H^{(370)}} = \Psi_{ML}$ & 0 & 0 & 0 & 0 & 0 & 759 & 12\,144 & 91\,080 & 425\,040 & 1313116 & $\binom{24}{w}$ \\
		\bottomrule
	\end{tabular}
	\renewcommand{\arraystretch}{1}% Back to normal
\end{table*}

Further, let $\Psi(w)$ be a number of erasure patterns of weight $w$, $0 \leq w \leq n$, in a code of length $n$, that cannot be decoded by some decoding method. Then, the frame error rate (also known as the block error rate) is a function of the bit erasure probability $p$, as follows:
\[
\operatorname{FER}(p) = \sum_{w=0}^{n} \Psi(w) p^w (1-p)^{n-w} .
\]

\begin{figure*}
	\centering
	\begin{subfloat}%[t]{0.5\textwidth}
		\centering
		\includegraphics[scale=.9]{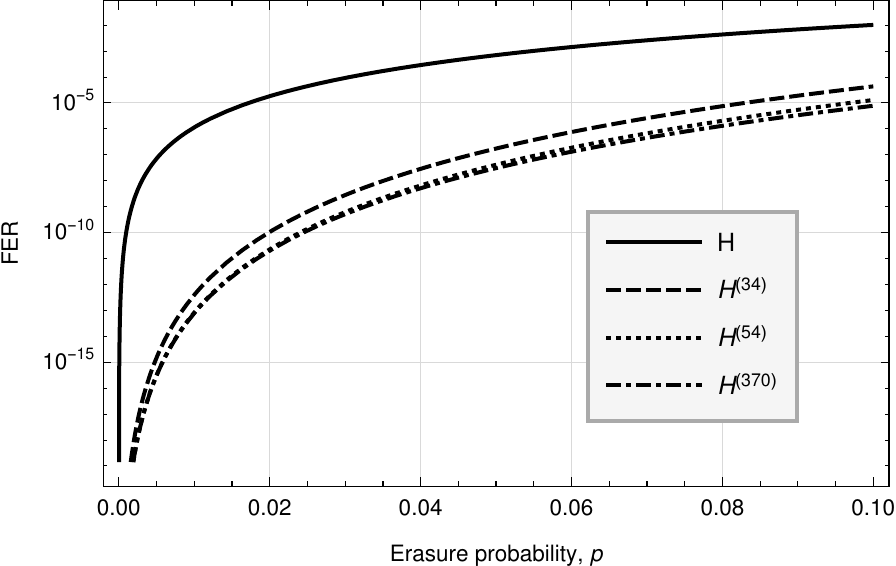}
	\end{subfloat}
	\begin{subfloat}%[t]{0.5\textwidth}
		\centering
		\includegraphics[scale=.9]{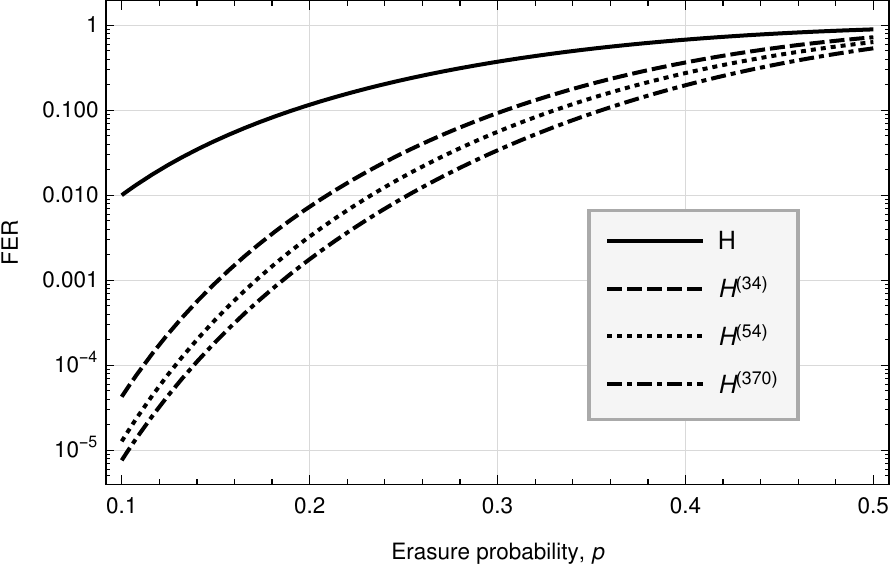}
	\end{subfloat}
	
	\caption{Frame error rates for different parity-check matrices of the extended Golay code, obtained by a randomized greedy algorithm. There are no coverable stopping sets of size up to 3, 7, 8, and 12 for $H$, $H^{(34)}$, $H^{(54)}$, and $H^{(370)}$, respectively.}
	\label{fig:Golay-SV-FERs}
\end{figure*}

Based on the number of undecodable erasure patterns, we plot the performance curves in \cref{fig:Golay-SV-FERs}. We note that plots for $H^{(54)}$ and larger matrices are almost visually indistinguishable from the plot for $H^{(370)}$. 

\begin{table*}[!t]
	\centering
	\caption{ML stopping redundancies average over $\mathfrak S(n,m)$. Estimates hold with probability 95\%}
	\label{tbl:SRE-simulations}
	
	\renewcommand{\arraystretch}{1.4}
	\begin{tabular}{@{}rrrrrcrrrcrrr@{}}
		\toprule
		
		\multirow{2}{*}{$n$} & \phantom{\small .} & \multicolumn{3}{c}{$R = \nicefrac 13$} & \phantom{\small .} & \multicolumn{3}{c}{$R = \nicefrac 12$} & \phantom{\small .} & \multicolumn{3}{c}{$R = \nicefrac 23$} \\
		\cmidrule{3-5} \cmidrule{7-9} \cmidrule{11-13}
		&& $\rho_m$ & $\hat \rho_m$ & $\epsilon^{(m)}$, \% && $\rho_m$ & $\hat \rho_m$ & $\epsilon^{(m)}$, \% && $\rho_m$ & $\hat \rho_m$ & $\epsilon^{(m)}$, \% \\
		
		\midrule
		6 && 6 & 6 & 1.27 && 3 & 3 & 1.7 && 2 & 2 & 2.53 \\
		12 && 84.99 & 85 & 0.64 && 34.75 & 34.77 & 0.85 && 10.55 & 10.55 & 1.27 \\
		18 && \np{1223.92} & \np{1224.18} & 0.43 && 281.32 & 281.37 & 0.57 && 46.11 & 46.12 & 0.85 \\
		24 && \np{18557} & \np{18557.6} & 0.32 && \np{2234.5} & \np{2234.82} & 0.43 && 189.07 & 189.08 & 0.64 \\
		30 && \np{288386} & \np{288422} & 0.26 && \np{17715.6} & \np{17717.1} & 0.34 && 758.87 & 758.9 & 0.51 \\
		36 && \np{4.5288e6} & \np{4.5301e6} & 0.21 && \np{140636} & \np{140645} & 0.28 && \np{3027.58} & \np{3027.7} & 0.43 \\
		42 && \np{7.1464e7} & \np{7.1467e7} & 0.18 && \np{1.1180e6} & \np{1.1181e6} & 0.24 && \np{12064.5} & \np{12065.1} & 0.37 \\
		48 && \np{1.1308e9} & \np{1.1310e9} & 0.16 && \np{8.8982e6} & \np{8.8987e6} & 0.21 && \np{48084} & \np{48085.4} & 0.32 \\
		54 && \np{1.7926e10} & \np{1.7928e10} & 0.14 && \np{7.0879e7} & \np{7.0883e7} & 0.19 && \np{191731} & \np{191734} & 0.28 \\
		\bottomrule
	\end{tabular}
	\renewcommand{\arraystretch}{1}
%	\begin{tabular}{c|ccc|ccc|ccc}
%		\toprule
%		%		\renewcommand{\arraystretch}{2}
%		\multirow{2}{*}{$n$} & \multicolumn{3}{c|}{$R = \nicefrac 13$} & \multicolumn{3}{c|}{$R = \nicefrac 12$} & \multicolumn{3}{c}{$R = \nicefrac 23$} \\
%		%		\cline{2-10}
%		& $\rho_m$ & $\hat \rho_m$ & $\epsilon^{(m)}$, \% & $\rho_m$ & $\hat \rho_m$ & $\epsilon^{(m)}$, \% & $\rho_m$ & $\hat \rho_m$ & $\epsilon^{(m)}$, \% \\
%		\midrule
%		6 & 6 & 6 & 1.27 & 3 & 3 & 1.7 & 2 & 2 & 2.53 \\
%		12 & 84.99 & 85 & 0.64 & 34.75 & 34.77 & 0.85 & 10.55 & 10.55 & 1.27 \\
%		18 & 1223.92 & 1224.18 & 0.43 & 281.32 & 281.37 & 0.57 & 46.11 & 46.12 & 0.85 \\
%		24 & 18557 & 18557.6 & 0.32 & 2234.5 & 2234.82 & 0.43 & 189.07 & 189.08 & 0.64 \\
%		30 & 288386 & 288422 & 0.26 & 17715.6 & 17717.1 & 0.34 & 758.87 & 758.9 & 0.51 \\
%		36 & $4.52875\cdot} 10^6$ & $4.5301\cdot} 10^6$ & 0.21 & 140636 & 140645 & 0.28 & 3027.58 & 3027.7 & 0.43 \\
%		42 & $7.14642\cdot} 10^7$ & $7.14672\cdot} 10^7$ & 0.18 & $1.11803\cdot} 10^6$ & $1.11805\cdot} 10^6$ & 0.24 & 12064.5 & 12065.1 & 0.37 \\
%		48 & $1.13081\cdot} 10^9$ & $1.13099\cdot} 10^9$ & 0.16 & $8.89819\cdot} 10^6$ & $8.89873\cdot} 10^6$ & 0.21 & 48084 & 48085.4 & 0.32 \\
%		54 & $1.79263\cdot} 10^{10}$ & $1.79282\cdot} 10^{10}$ & 0.14 & $7.08789\cdot} 10^7$ & $7.08828\cdot} 10^7$ & 0.19 & 191731 & 191734 & 0.28 \\
%		\bottomrule
%	\end{tabular}
\end{table*}

\subsection{Standard random ensemble}
In this section, we apply the results of \cref{lem:conf-interval-ens} to the standard random ensemble $\mathfrak S(n,m)$. We calculate estimates on $\Expect[\mathfrak S(n,m)]{ \rho_{\ell}(\code) }$ for different $n$ and $m = (1-R)n$ for ``design'' code rates $R \in \{ \nicefrac 13, \nicefrac 12, \nicefrac 23 \}$. For each pair $(n, m)$ and each size $i=1,2,\dotsc,m,$ we generate $N = 10^7$ pairs  $\left( H^{(i)}, \cS^{(i)} \right)$ and register the frequencies of $\cS^{(i)}$ being a coverable stopping set in $H^{(i)}$.

Based on the frequencies, we obtain estimates $\hat{\bar u}_i$ on the ensemble-average sizes $\bar u_i$. For each size of the stopping sets $i$, we use $\epsilon_i = 1 - 0.95^{1/m}$, which gives a confidence of $95\%$ that the estimates on $\bar u_i$ hold.

After that, we apply \cref{cor:rho-ens-average} in order to obtain bounds on $\Expect[\ensemble]{ \rho_{m}(\code)}$, for selected values of $m$. These bounds are denoted by $\hat \rho_m$. \cref{tbl:SRE-simulations} presents the resulting values. 
They are compared to the values $\Xi_{\ell}^{(II)} (\bar{u}_1, \bar{u}_2, \dotsc, \bar{u}_{\ell})$ (obtained analytically, and denoted by $\rho_m$). We observe that the numerical results are a very good approximation to the theoretical values.

\subsection{Gallager ensemble}\label{sec:Gallager-average}
We repeat the experiments of the previous subsection on the Gallager ensemble $\mathfrak{Gal}(n, J, K)$ of $(J,K)$-regular LDPC codes~\cite{gallager1962low} for different choices of $(J,K)$ and different lengths $n$. The ensemble is defined by parity-check matrices of a special form. An $\left( \frac{n J}K \right) \times n$ parity-check matrix consists of $J$ strips of width $M = n/K$ rows each. In the first strip, the $j$th row contains $K$ ones in positions $(j-1)K+1, (j-1)K+2, \dotsc, jK$ for $j=1,2,\dotsc,M$. Each of the other strips is a random column permutation of the first strip.

It is known that the rank of a parity-check matrix in $\mathfrak{Gal}(n,J,K)$ cannot be larger than $r_{\rm max} = \nicefrac{n J}{K} - (J-1)$ due to the presence of redundant rows in any such matrix. Therefore, the ML decoding performance is achieved when all the coverable stopping sets of size up to $r_{\rm max}$ are covered.

\begin{figure}
	\begin{center}
		\includegraphics[scale=.7]{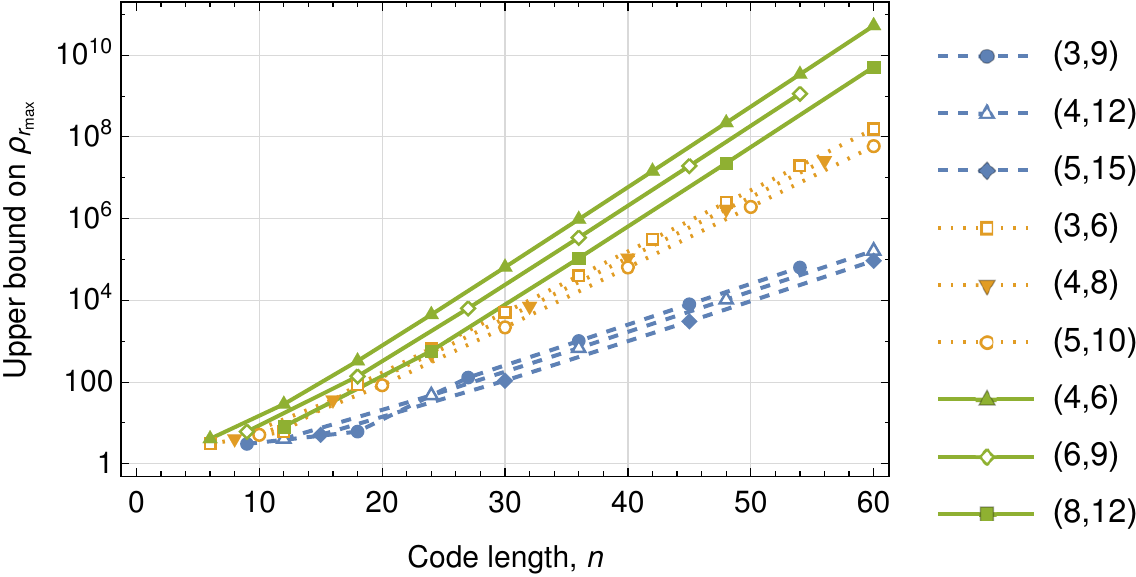}
		\caption{Upper bounds on $\mathfrak{Gal}(n, J, K)$-average $r_{\rm max}$-th stopping redundancy}
		\label{fig:GAL-plot}
	\end{center}
\end{figure}

\cref{fig:GAL-plot} demonstrates the values of the ML stopping redundancy, $\rho_{r_{\rm max}}$, for different lengths and different choices of $J$ and $K$. We observe three clusters of plots according to the design rates of the codes.

\section{Conclusions}
\label{sec:conclusions}
In this work, we observed that 
stopping sets of size $\ell$, $d \le \ell \le r$, are important for analysis of ML decoding over the BEC. We generalized the analytical approach in~\cite{schwartz-vardy2006} by estimating the number of stopping sets of such sizes. This novel approach led to estimates on the number of redundant rows in the parity-check matrix needed in order to make the performance of the iterative decoder similar to that of ML decoding. The results were tested numerically, and the experimental results were compared to the theoretical counterparts. 

%\lipsum[1]

% if have a single appendix:
%\appendix[Proof of the Zonklar Equations]
% or
%\appendix  % for no appendix heading
% do not use \section anymore after \appendix, only \section*
% is possibly needed

% use appendices with more than one appendix
% then use \section to start each appendix
% you must declare a \section before using any
% \subsection or using \label (\appendices by itself
% starts a section numbered zero.)
%

\appendices
\crefalias{section}{appsec}
\section{}\label{app:binoms-divide-lemma}
This appendix explains in details the equality \cref{eq:binoms-divide-for-lemma} used in the proof of \cref{thm:sr-itw}.

\begin{lemma}\label{lem:frac-binoms}
	For all non-negative integers $r$, $i$, $\tau$, and $t$ such that $0 \leq t \leq (2^r - \tau - 1) - i \cdot 2^{r-i}$,
	\ifonecolumn
	\begin{equation*}
	\left.
	\binom{(2^r - \tau - 1) - i \cdot 2^{r-i}}{t}
	\right/
	{\binom{2^r - \tau - 1}{t}}
	= \prod_{j=\tau+1}^{\tau+t} \pi(r,i,j).
	\end{equation*}
	\else
	\begin{multline*}
		\left.
		\binom{(2^r - \tau - 1) - i \cdot 2^{r-i}}{t}
		\right/
		{\binom{2^r - \tau - 1}{t}}\\
		= \prod_{j=\tau+1}^{\tau+t} \pi(r,i,j).
	\end{multline*}
	\fi
\end{lemma}
\begin{IEEEproof}
	Indeed,
	\ifonecolumn
	\begin{align*}
	\left.
	\binom{(2^r - \tau - 1) - i \cdot 2^{r-i}}{t}
	\right/
	{\binom{2^r - \tau - 1}{t}} &= \frac{\left( 2^r-\tau-1-i\cdot 2^{r-i} \right)!}{t! \left( 2^r-\tau-1-i\cdot 2^{r-i}-t \right)!}
	\cdot
	\frac{t! \left( 2^r-\tau-1-t \right)!}{\left( 2^r-\tau-1 \right)!}\\
	&= \prod_{j=\tau+1}^{\tau+t} \frac{2^r-j-i\cdot 2^{r-i}}{2^r-j}
	= \prod_{j=\tau+1}^{\tau+t} \pi(r,i,j).\qedhere
	\end{align*}
	\else
	\begin{align*}
		&\left.
		\binom{(2^r - \tau - 1) - i \cdot 2^{r-i}}{t}
		\right/
		{\binom{2^r - \tau - 1}{t}} \\
		&\quad= \frac{\left( 2^r-\tau-1-i\cdot 2^{r-i} \right)!}{t! \left( 2^r-\tau-1-i\cdot 2^{r-i}-t \right)!}
		\cdot
	 	\frac{t! \left( 2^r-\tau-1-t \right)!}{\left( 2^r-\tau-1 \right)!}\\
		&\quad= \prod_{j=\tau+1}^{\tau+t} \frac{2^r-j-i\cdot 2^{r-i}}{2^r-j}
		= \prod_{j=\tau+1}^{\tau+t} \pi(r,i,j).\qedhere
	\end{align*}
	\fi
\end{IEEEproof}

\section{}\label{app:w-opt}
In this appendix, we aim to find a weight $w$ of a row in a parity-check matrix, which covers the maximum number of stopping sets of size up to $\ell$, provided that $n$ is fixed. It is easy to see that any row of length $n$ and  weight $w$ covers exactly
\[
	w \sum_{i=1}^{\ell} \binom{n-w}{i-1}
\]
stopping sets of weight up to $\ell$. \Cref{lem:w-opt} provides an answer to this maximization question.

\begin{lemma}\label{lem:w-opt}
	Fix two positive integers $n$ and $2 \leq \ell \leq n$ and define a discrete function $F \colon \{ 1, 2, \dotsc, n-\ell+1\} \rightarrow \bbN$ in the following way:
	\[
	F(w) = F_{n,\ell}(w) = w \sum_{i=0}^{\ell-1} \binom{n-w}{i} .
	\]
	Then
	\[
	\arg \max_w F(w) \in \left\{ \left\lfloor \frac{n+1}{\ell} \right\rfloor, \left\lceil \frac{n}{\ell} \right\rceil \right\}   .
	\]
\end{lemma}
\begin{IEEEproof}
	In order to prove the statement of the lemma, it is sufficient to show that $F(w)$ increases for $w \leq \left\lfloor \frac{n+1}{\ell} \right\rfloor$ and decreases for $w \geq \left\lceil \frac{n}{\ell} \right\rceil$.
		
	Consider the finite difference:
	\[
		\Delta F(w) = F(w+1) - F(w) .
	\]
	
	It can be expanded as follows:
	\ifonecolumn
	\begin{equation*}
	\Delta F(w) = F(w+1) - F(w)
	= (w+1)\sum_{i=0}^{\ell-1}\binom{n-w-1}{i} - w\sum_{i=0}^{\ell-1}\binom{n-w}{i}.
	\end{equation*}
	\else
	\begin{multline*}
		\Delta F(w) = F(w+1) - F(w)\\
		 = (w+1)\sum_{i=0}^{\ell-1}\binom{n-w-1}{i} - w\sum_{i=0}^{\ell-1}\binom{n-w}{i}.
	\end{multline*}
	\fi

	Consequently,
	\ifonecolumn
	\else
	\begin{align*}
		\Delta F(w) &= (w+1)\sum_{i=0}^{\ell-1}\binom{n-w-1}{i}\\
		&\quad- w\sum_{i=0}^{\ell-1}\left(\binom{n-w-1}{i} + \binom{n-w-1}{i-1}\right)\\
		&= \sum_{i=0}^{\ell-1}\binom{n-w-1}{i} - w\sum_{i=0}^{\ell-1}\binom{n-w-1}{i-1}.
	\end{align*}
	\fi
			
	We have:
	\ifonecolumn
	\begin{equation*}
	\Delta F(w) = \sum_{i=0}^{\ell-1}\left( \binom{n-w-1}{i} - w\binom{n-w-1}{i-1} \right)
	= \sum_{i=0}^{\ell-1} \frac{(n-w-1)!}{i!(n-w-i)!}\left( n-i-w(i+1) \right).
	\end{equation*}
	\else
	\begin{align*}
		\Delta F(w) &= \sum_{i=0}^{\ell-1}\left( \binom{n-w-1}{i} - w\binom{n-w-1}{i-1} \right)\\
		&= \sum_{i=0}^{\ell-1} \frac{(n-w-1)!}{i!(n-w-i)!}\left( n-i-w(i+1) \right).
	\end{align*}
	\fi
	If we require that 
	\[
	w \leq \frac{n-\ell+1}{\ell} ,
	\]
	then it follows also that
	\[
	w < \frac{n-i}{i+1} \quad \text{ for all } i < \ell-1.
	\]
	Hence, each of the terms $(n-i-w(i+1))$ is positive for $i < \ell-1$ and $(n-\ell+1 - w\ell) \geq 0$. Therefore,
	\[
	F(1) < F(2) < \dotsb %< F\left(\left\lfloor \frac{n+1}{\ell} - 1 \right\rfloor\right) 
	< F\left(\left\lfloor \frac{n+1}{\ell} \right\rfloor\right) .
	\]
	
	On the other hand, we can write:
	\ifonecolumn
	\begin{align*}
	\Delta F(w) = \sum_{i=0}^{\ell-1}\binom{n-w-1}{i} - w\sum_{i=0}^{\ell-1}\binom{n-w-1}{i-1}
	&= \sum_{i=0}^{\ell-1}\binom{n-w-1}{i} - w\sum_{i=0}^{\ell-2}\binom{n-w-1}{i}\\
	&= \binom{n-w-1}{\ell-1} + (1-w)\sum_{i=0}^{\ell-2}\binom{n-w-1}{i}.
	\end{align*}
	\else
	\begin{align*}
		\Delta F(w) &= \sum_{i=0}^{\ell-1}\binom{n-w-1}{i} - w\sum_{i=0}^{\ell-1}\binom{n-w-1}{i-1}\\
		&= \sum_{i=0}^{\ell-1}\binom{n-w-1}{i} - w\sum_{i=0}^{\ell-2}\binom{n-w-1}{i}\\
		&= \binom{n-w-1}{\ell-1} + (1-w)\sum_{i=0}^{\ell-2}\binom{n-w-1}{i}.
	\end{align*}
	\fi
	And, if $w > 1$, we have:
	\ifonecolumn
	\begin{equation*}
	\Delta F(w) < \binom{n-w-1}{\ell-1} + (1-w)\binom{n-w-1}{\ell-2}
	= \frac{(n-w-1)!}{(\ell-1)!(n-\ell-w+1)!} (n-w\ell) .
	\end{equation*}
	\else
	\begin{align*}
		\Delta F(w) &< \binom{n-w-1}{\ell-1} + (1-w)\binom{n-w-1}{\ell-2}\\
		&= \frac{(n-w-1)!}{(\ell-1)!(n-\ell-w+1)!} (n-w\ell) .
	\end{align*}
	\fi
			
	If we further require $w \ge \frac{n}{\ell}$, then $\Delta F(w) < 0$ and
	\[
		F\left( \left\lceil \frac{n}{\ell} \right\rceil \right) > F\left( \left\lceil \frac{n}{\ell} \right\rceil +1 \right) >  \dots > F(n-\ell+1).\qedhere
	\]
\end{IEEEproof}

%%%%%%%%%%%%%%%%%%%%%%%%%%%%%%%%%%%%%%%%%%%%%%%%%%%%%%%%%%%%%%%%%%%%%%%%%%%%%%%%%%%%%%%%%

\section{}\label{app:alekseyev}
In this appendix, we compute the number of full-rank binary matrices with no rows of weight one.
The results in this appendix are based on~\cite{alekseyev}.

\begin{lemma}\label{lem:alekseyev-number}
	Let $m \geq i$ and denote by $\mathcal N(m, i)$ the number of full-rank binary $m \times i$ matrices with no rows of Hamming weight one. Then
	\ifonecolumn
	\begin{equation}
	\mathcal N(m, i) = 
	\sum_{k=0}^i \binom{i}{k}\cdot k! \sum_{p=0}^{m} (-1)^{m-p}\cdot \binom{m}{p}
	\cdot 2^{kp}\cdot S(m-p,k) \prod_{t=0}^{i-k-1} (2^p - 2^t) ,
	\label{eq:N}
	\end{equation}
	\else
	\begin{multline}
	\mathcal N(m, i) = 
	\sum_{k=0}^i \binom{i}{k}\cdot k! \sum_{p=0}^{m} (-1)^{m-p}\cdot \binom{m}{p}\\
	\cdot 2^{kp}\cdot S(m-p,k) \prod_{t=0}^{i-k-1} (2^p - 2^t) ,
	\label{eq:N}
	\end{multline}
	\fi
	where $S(x,y)$ is a Stirling number of the second kind (the number of ways to partition a set of $x$ labelled objects into $y$ nonempty unlabelled subsets).
\end{lemma}
\begin{IEEEproof}
	First, we calculate the number of $m \times i$ binary matrices of full rank, which we denote by $\mathcal M(m, i)$. As $m \geq i$, all the columns in such matrices are linearly independent. We have $2^m - 1$ choices for the first column (any nonzero vector in $\mathbb F_2^m$), $2^m - 2$ choices for the second column (any vector in $\mathbb F_2^m$ except the all-zero vector and the first column), $2^m - 2^2$ choices for the third column (any vector in $\mathbb F_2^m$ except for the vectors in the subspace spanned by the first two columns), etc. Altogether, we have
	\[
	\mathcal M(m, i) = \prod_{t=0}^{i-1} (2^m - 2^t) .
	\]
	
	Next, the number of full-rank $m \times i$ matrices with exactly $z$ zero rows can be obtained by using the inclusion-exclusion principle, as follows:
	\begin{equation}\label{eq:alexeyev-exactly-z-zeros}
	\binom mz \sum_{p=0}^{m-z} (-1)^{m-z-p} \binom{m-z}{p} \prod_{t=0}^{i-1} (2^p - 2^t) .
	\end{equation}
	
	Now, let us consider the requirement not to have rows of weight one. We use the inclusion-exclusion principle.
	
	Let $P_\iota$ ($\iota=1,2,\dotsc,i$) be the property that there is a row with a single 1 at $\iota$'th coordinate. Suppose that an $m \times i$ matrix satisfies properties with indices from a set $R \subseteq [i]$ with $|R| = k$. Then the set of row indices is partitioned as 
	\[
		[m] = J \sqcup \bar J ,
	\]
	where $J$ consists of indices corresponding to rows with a single 1 at a coordinate from $R$, and $\bar J = [m] \setminus J$. Let $|J| = j$ (we have $j \geq k$).
	
	To enumerate possible submatrices, whose rows are indexed by $J$ and columns by $[i]$, we notice that their columns essentially define an ordered partition of their rows into $k$ nonempty sets. Hence, the number of such submatrices equals to $k! \cdot S(j,k)$.
	
	The number of submatrices whose rows and columns are indexed by $\bar J$ and $\bar R$, respectively, with exactly $z$ zero rows can be calculated from \cref{eq:alexeyev-exactly-z-zeros}. They can be extended to all submatrices with rows indexed by $\bar J$ in $(2^k - k)^z(2^k)^{m-j-z}$ ways because each zero row can be extended by anything except for $k$-vectors of weight 1 (as we already collected them in rows $J$), and others can be extended by anything.
	
	Putting all together, we have
	\ifonecolumn
	\begin{align*}
	\mathcal N(m, i)  = 
	&\sum_{k=0}^i (-1)^k \binom{i}{k} \sum_{j=k}^m \binom{m}{j}\cdot k!\cdot S(j,k) \cdot \sum_{z=0}^{m-j} \binom{m-j}{z}\cdot (2^k-k)^z\cdot (2^k)^{m-j-z}\\
	&\cdot \sum_{p=0}^{m-j-z} (-1)^{m-j-z-p} \binom{m-j-z}{p} \cdot \prod_{t=0}^{i-k-1} (2^p - 2^t).
	\end{align*}
	\else
	\begin{align*}
	\mathcal N(m, i)  = 
	&\sum_{k=0}^i (-1)^k \binom{i}{k} \sum_{j=k}^m \binom{m}{j}\cdot k!\cdot S(j,k)\\
	&\cdot \sum_{z=0}^{m-j} \binom{m-j}{z}\cdot (2^k-k)^z\cdot (2^k)^{m-j-z}\\
	&\cdot \sum_{p=0}^{m-j-z} (-1)^{m-j-z-p} \binom{m-j-z}{p}\\
	&\cdot \prod_{t=0}^{i-k-1} (2^p - 2^t).
	\end{align*}
	\fi
	This can be further rewritten as
	\ifonecolumn
	\begin{equation*}
	\mathcal N(m, i)  = 
	\sum_{k=0}^i (-1)^k \binom{i}{k} \sum_{j=k}^m \binom{m}{j}\cdot k! \cdot S(j,k) \cdot \sum_{p=0}^{m-j} (-1)^{m-j-p} \binom{m-j}{p} \cdot 2^{kp}\cdot k^{m-j-p} \prod_{t=0}^{i-k-1} (2^p - 2^t),
	\end{equation*}
	\else
	\begin{align*}
	\mathcal N(m, i)  = 
	&\sum_{k=0}^i (-1)^k \binom{i}{k} \sum_{j=k}^m \binom{m}{j}\cdot k!\\
	&\cdot S(j,k) \cdot \sum_{p=0}^{m-j} (-1)^{m-j-p} \binom{m-j}{p}\\
	&\cdot 2^{kp}\cdot k^{m-j-p} \prod_{t=0}^{i-k-1} (2^p - 2^t),
	\end{align*}
	\fi
	which is in turn equivalent to
	\ifonecolumn
	\begin{align*}
	\mathcal N(m, i) &= \sum_{k=0}^i \binom{i}{k} \sum_{\ell=0}^k (-1)^{k-\ell}\cdot \binom{k}{\ell}
	\cdot \sum_{p=0}^{m} \binom{m}{p} \cdot 2^{kp}\cdot (-\ell)^{m-p} \prod_{t=0}^{i-k-1} (2^p - 2^t)\\
	&= \sum_{k=0}^i \binom{i}{k}\cdot k! \sum_{p=0}^{m} (-1)^{m-p}\cdot \binom{m}{p} \cdot 2^{kp}\cdot S(m-p,k) \prod_{t=0}^{i-k-1} (2^p - 2^t).\qedhere
	\end{align*}
	\else
	\begin{align*}
	\mathcal N(m, i) &= \sum_{k=0}^i \binom{i}{k} \sum_{\ell=0}^k (-1)^{k-\ell}\cdot \binom{k}{\ell}
	\cdot \sum_{p=0}^{m} \binom{m}{p}\\
	&\quad \cdot 2^{kp}\cdot (-\ell)^{m-p} \prod_{t=0}^{i-k-1} (2^p - 2^t)\\
	&= \sum_{k=0}^i \binom{i}{k}\cdot k! \sum_{p=0}^{m} (-1)^{m-p}\cdot \binom{m}{p}\\
	&\quad \cdot 2^{kp}\cdot S(m-p,k) \prod_{t=0}^{i-k-1} (2^p - 2^t).\qedhere
	\end{align*}
	\fi
\end{IEEEproof}

We note that for the medium and large values of $m$ and $i$, the ratio of the number of full-rank binary $m \times i$ matrices without rows of weight one to the number of all full-rank binary matrices is very close to $1$, and hence the relative error becomes close to $0$. For example, for $m = 50$ and $i = 30$ we have
\[
	\frac{\mathcal M(50, 30) - \mathcal N(50, 30)}{\mathcal N(50, 30)} \approx 1.40 \times 10^{-6}.
\]
Since obviously $\mathcal M(m,i) \geq \mathcal N(m, i)$, the former is a correct upper bound, which is rather tight for the medium and large values of $m$ and $i$. For practical purposes, it is much easier to calculate and analyze $\mathcal M(m,i)$ than $\mathcal N(m,i)$.

% use section* for acknowledgment
\section*{Acknowledgment}
The authors would like to thank \texttt{mathoverflow.net} community and especially Max Alekseyev for help with the proof of \cref{lem:alekseyev-number}.

% Can use something like this to put references on a page
% by themselves when using endfloat and the captionsoff option.
\ifCLASSOPTIONcaptionsoff
  \newpage
\fi

% trigger a \newpage just before the given reference
% number - used to balance the columns on the last page
% adjust value as needed - may need to be readjusted if
% the document is modified later
%\IEEEtriggeratref{8}
% The "triggered" command can be changed if desired:
%\IEEEtriggercmd{\enlargethispage{-5in}}

% references section

%\nocite{*}

% can use a bibliography generated by BibTeX as a .bbl file
% BibTeX documentation can be easily obtained at:
% http://mirror.ctan.org/biblio/bibtex/contrib/doc/
% The IEEEtran BibTeX style support page is at:
% http://www.michaelshell.org/tex/ieeetran/bibtex/
\bibliographystyle{IEEEtran}
% argument is your BibTeX string definitions and bibliography database(s)
\bibliography{IEEEabrv,stopredjrnl}
%
% <OR> manually copy in the resultant .bbl file
% set second argument of \begin to the number of references
% (used to reserve space for the reference number labels box)
%\begin{thebibliography}{1}
%
%\bibitem{IEEEhowto:kopka}
%H.~Kopka and P.~W. Daly, \emph{A Guide to \LaTeX}, 3rd~ed.\hskip 1em plus
%  0.5em minus 0.4em\relax Harlow, England: Addison-Wesley, 1999.
%
%\end{thebibliography}

% biography section
% 
% If you have an EPS/PDF photo (graphicx package needed) extra braces are
% needed around the contents of the optional argument to biography to prevent
% the LaTeX parser from getting confused when it sees the complicated
% \includegraphics command within an optional argument. (You could create
% your own custom macro containing the \includegraphics command to make things
% simpler here.)
%\begin{IEEEbiography}[{\includegraphics[width=1in,height=1.25in,clip,keepaspectratio]{mshell}}]{Michael Shell}

% if you will not have a photo at all:
\begin{IEEEbiographynophoto}{Yauhen Yakimenka} was born in Baranavichy, Belarus, in 1986. He received the B.Sc. degree in computer security from Belarusian State University, Minsk, Belarus, in 2008, and M.Sc. joint degree in cyber security from Tallinn University of Technology, Estonia, and University of Tartu, Estonia, in 2014. Since then, he has been working on  his Ph.D. degree at the Institute of Computer Science, University of Tartu.
\end{IEEEbiographynophoto}

\begin{IEEEbiographynophoto}{Vitaly Skachek} received the B.A. (Cum Laude), M.Sc. and Ph.D. degrees in computer science from the Technion---Israel Institute of Technology, in 1994, 1998 and 2007, respectively.

Since 2012, Dr. Skachek is employed by the Institute of Computer Science, University of Tartu, where he is an Associate Professor. He held visiting positions with the Mathematics of Communications Department, Bell Laboratories, Murray Hill, with the Claude Shannon Institute, University College Dublin, Dublin, with the School of Physical and Mathematical Sciences, Nanyang Technological University, Singapore, with the Coordinated Science Laboratory, University of Illinois at Urbana-Champaign, Urbana, and with the Department of Electrical and Computer Engineering, McGill University, Montreal. He is a member of the editorial board for the \emph{Advances in Mathematics of Communications} and a guest editor for the special issue of the \emph{Cryptography and Communications -- Discrete Structures, Boolean Functions and Sequences}. He served as a guest editor for the special issue on network coding of the \emph{EURASIP Journal on Advances in Signal Processing}.
	
Dr. Skachek is a recipient of the Permanent Excellent Faculty Instructor award, given by Technion.
\end{IEEEbiographynophoto}

\begin{IEEEbiographynophoto}{Irina E. Bocharova} received the Diploma in Electrical Engineering in 1978 from the Leningrad Electro-technical Institute and the Ph.D. degree in technical sciences in 1986 from Leningrad Institute of Aircraft Instrumentation. Since 1986 until 2007, she has been Senior Researcher, Assistant Professor, and then Associate Professor at the Leningrad Institute of Aircraft Instrumentation (now State University of Aerospace Instrumentation, St.~Petersburg, Russia). Since 2007 she has been Associate Professor at the St.~Petersburg State University of Information Technologies, Mechanics and Optics. Additionally, since 2016 she has been Senior Research Fellow of Coding Theory at the Institute of Computer Science of the University of Tartu.
	
Her research interests include convolutional codes, communication systems, source coding and its applications to speech, audio and image coding. She has more than 70 papers published in journals and proceedings of international conferences, and seven U.S. patents in speech, audio and video coding. She has authored the textbook \emph{Compression for Multimedia} (Cambridge University Press, 2010). Professor Bocharova was awarded the Lise Meitner Visiting Chair in engineering at Lund University, Lund, Sweden twice (January--June 2005 and 2011). In 2014 she received Scopus Russia Award.
\end{IEEEbiographynophoto}

\begin{IEEEbiographynophoto}{Boris D. Kudryashov} received the Diploma in Electrical Engineering in 1974 and the Ph.D. in technical sciences degree in 1978 both from the Leningrad Institute of Aerospace Instrumentation, and the Doctor of Science degree in 2005 	from Institute of Problems of Information Transmission, Moscow. In 1978, he became Assistant Professor, in 1983 Associate Professor, and in 2005 Professor at Leningrad Institute of Aerospace Instrumentation (now the State University on Aerospace Instrumentation, St.-Petersburg, Russia). Since November 2007, he has been Professor at the State University on Information Technologies, Mechanics and Optics, St.~Petersburg, Russia. Additionally, since 2017 he has been Visiting Associate Professor at the University of Tartu.

His research interests include coding theory, information theory and applications to speech, audio and image coding. He has authored two textbooks on information theory and coding theory (both in Russian), and has more than 80 papers published in journals and proceedings of international conferences, 20 US patents in image, speech and audio coding. Professor Kudryashov served as a member of the Organizing Committees of ACCT International Workshops.
\end{IEEEbiographynophoto}

\balance

%\begin{IEEEbiographynophoto}{Yauhen Yakimenka}
%Biography text here.
%\end{IEEEbiographynophoto}

%\begin{IEEEbiographynophoto}{Vitaly Skachek}
%Biography text here.
%\end{IEEEbiographynophoto}

% insert where needed to balance the two columns on the last page with
% biographies
%\newpage

%\begin{IEEEbiographynophoto}{Irina E. Bocharova}
%Biography text here.
%\end{IEEEbiographynophoto}

%\begin{IEEEbiographynophoto}{Boris D. Kudryashov}
%	Biography text here.
%\end{IEEEbiographynophoto}

% You can push biographies down or up by placing
% a \vfill before or after them. The appropriate
% use of \vfill depends on what kind of text is
% on the last page and whether or not the columns
% are being equalized.

%\vfill

% Can be used to pull up biographies so that the bottom of the last one
% is flush with the other column.
%\enlargethispage{-5in}

% that's all folks
\end{document}